\documentclass[12pt]{article}

\usepackage[T1]{fontenc}
\usepackage[section]{placeins}
\usepackage{rotating}
\usepackage{color}
\usepackage[english]{babel}
\usepackage{standalone}
\usepackage{float}
\usepackage{booktabs}
\usepackage{rotating}
\usepackage{graphics}
\graphicspath{ {Fig/} }
\usepackage{listings}
\usepackage{setspace}
\usepackage{natbib}
\bibpunct{(}{)}{;}{a}{,}{,}
\usepackage{url}
\usepackage[titletoc,title]{appendix}
\usepackage[left=1.5cm,right=1.5cm,top=2cm,bottom=2cm]{geometry}
\usepackage[font=footnotesize, labelfont={bf}, margin=1cm]{caption}
\usepackage{rotating}
\usepackage{array}
\usepackage{colortbl}
\usepackage{enumerate,multicol}
\usepackage{amsmath, amssymb, amsthm}
\usepackage{enumitem,bm}
\usepackage{parcolumns}

\providecommand{\keywords}[1]{{\bf{Key words:}} #1}

\usepackage[multiple]{footmisc}
\usepackage{authblk}
\usepackage[position=bottom]{subfig}
\usepackage{multirow}
\usepackage{makecell}
\usepackage{fancyvrb} 
\usepackage{listings,color} 
\usepackage{moreverb}

\newenvironment{nalign}{
    \begin{equation}
    \begin{aligned}
}{
    \end{aligned}
    \end{equation}
    \ignorespacesafterend
}

\usepackage{pifont}
\newcommand{\cmark}{\ding{51}}%
\newcommand{\xmark}{\ding{55}}%
\newcommand\ack{\section*{Acknowledgement}}

\begin{document}

\title{\bf{Joint longitudinal models for dealing with missing at random data in trial-based economic evaluations}}
\author[1]{Andrea Gabrio\thanks{E-mail: andrea.gabrio.15@ucl.ac.uk}}
\author[2]{Rachael Hunter}
\author[3]{Alexina J. Mason}
\author[1]{Gianluca Baio}
\affil[1]{\small \textit{Department of Statistical Science, University College London,~UK}}
\affil[2]{\small \textit{Research Department of Primary Care and Population Health, University College London Medical School,~UK}}
\affil[3]{\small \textit{Department of Health Services Research and Policy, London School of Hygiene and Tropical Medicine,~UK}}

\date{}
\maketitle

\vspace*{-1cm}
\abstract{
Health economic evaluations based on patient-level data collected alongside clinical trials~(e.g.~health related quality of life and resource use measures) are an important component of the process which informs resource allocation decisions. Almost inevitably, the analysis is complicated by the fact that some individuals drop out from the study, which causes their data to be unobserved at some time point. Current practice performs the evaluation by handling the missing data at the level of aggregated variables (e.g.~QALYs), which are obtained by combining the economic data over the duration of the study, and are often conducted under a missing at random (MAR) assumption. However, this approach may lead to incorrect inferences since it ignores the longitudinal nature of the data and may end up discarding a considerable amount of observations from the analysis. We propose the use of joint longitudinal models to extend standard cost-effectiveness analysis methods by taking into account the longitudinal structure and incorporate all available data to improve the estimation of the targeted quantities under MAR. Our approach is compared to popular missingness approaches in trial-based analyses, motivated by an exploratory simulation study, and applied to data from two real case studies. 
}

\vspace*{0.5cm}
\vspace*{0.5cm}
\hrule

\clearpage

\section{Introduction}\label{intro}
Trial-based cost-effectiveness analyses (CEAs) rely on patient-level health economic data, which are collected at baseline and one or more follow-up points in the study through self-reported questionnaires, e.g.~the EuroQol (EQ)-5D~\citep{EQ5D}, and patient files. These data are combined with national value sets and unit prices to generate utility and cost measures at each time point in the study, and then aggregated over the study period into some patient-level effectiveness, e.g.~Quality-Adjusted Life Years (QALYs), and total cost outcomes, which represent the quantities of interest for the economic analysis.

A typical problem of trial-based CEAs is that some individuals fail to follow-up or are associated with some missing utility/cost values at some time point (thus impairing the calculation of their aggregated measures), which forces the analyst to make subjective assumptions about the unobserved data. These assumptions are subjective in that data cannot be used to critique them. \citet{Rubina} introduced three general classes of \textit{mechanisms} responsible for the missing values, which offer a convenient way to formulate assumptions about the reasons of missingness. The three mechanisms are: \textit{missing completely at random} (MCAR), where the probability of missingness is unrelated to observed or unobserved data; \textit{missing at random} (MAR), where the probability of missingness is unrelated to unobserved data conditional on observed data; and \textit{missing not at random} (MNAR), where the probability of missingness depends on unobserved data conditional on observed data. 

Different methods have been used to handle missing data in trial-based CEAs, each relying on different assumptions about the missingness mechanism. In routine analyses, a popular approach is given by \textit{case deletion} methods, which restrict the evaluation to the individuals with observed utility and cost data at each time point in the study~\citep{Noble2012,Gabrio,Leurent2018}. The model of interest is fitted to the complete cases and estimates are generated either only from the completers or from all available data, in which cases the methods take the name of \textit{complete case analysis} (CCA) and \textit{available case analysis} (ACA), respectively. Despite being simple to implement, both CCA and ACA are inefficient, lead valid inferences only under MCAR and contravene the intention to treat principle that all randomised patients should be included in the analysis~\citep{Little2002}. We note that, when the estimates of interest are generated through the incorporation of some baseline covariates (e.g.~using a regression approach), case deletion methods can lead inferences that are valid under less restrictive assumptions than MCAR, since they are defined conditional on the covariates included (an assumption referred to as covariate dependent~MCAR~\citep{little1995modeling}). In randomised controlled trials, \textit{baseline imputation} methods are an alternative approach which allows to retain the full sample by replacing the missing data in the baseline covariates with a single imputed value. Among these approaches, \textit{mean imputation} (MEAN) is the most commonly used, which replaces the missing data for each baseline covariate with the mean of the observed values across the intervention groups~\citep{White2005}. Baseline imputation methods are more efficient than CCA or ACA and can lead valid inferences under MAR~\citep{Sullivan2016}. However, in trial-based CEAs, a drawback of these methods is that the aggregated outcomes are not typically available for individuals with a missing baseline utility/cost value, therefore also requiring imputation for the outcome variables.  

Estimation methods based on a joint model for baseline and outcome data have been suggested to more appropriately assess the impact of missing data uncertainty on the estimates~\citep{Little1992}. In particular, two flexible approaches to perform joint modelling are \textit{multiple imputation} (MI) and \textit{full Bayesian} (FB) methods. MI defines an imputation model to produce multiple values for each missing observation, fits the model of interest to each imputed dataset, and uses Rubin's rules to combine the parameter estimates into a pooled estimate~\citep{Rubina}. Provided that the imputation model is correctly specified, MI gives consistent and asymptotically efficient parameter estimates under MAR~\citep{Carpenter}. Among the various approaches that are available, multiple imputation by chained equations (MICE) is one of the most popular among practitioners~\citep{VanBuuren}. In contrast to MI, which separates the imputation and analysis steps, FB methods estimate the parameters of interest simultaneously with the imputation of the missing values. FB methods are typically implemented using some iterative algorithms such as \textit{Mark Chain Monte Carlo} (MCMC)~\citep{Brooks} methods and allow the propagation of the uncertainty throughout each unobserved quantity in the model (either parameters or missing values). If weakly informative prior distributions are specified, inferences from FB are based on the observed data and are valid under~MAR~\citep{Daniels}.

\subsection{Standard Approach in Trial-Based CEAs}\label{long}
\begin{figure}[H]
\centering
\includegraphics[scale=0.7]{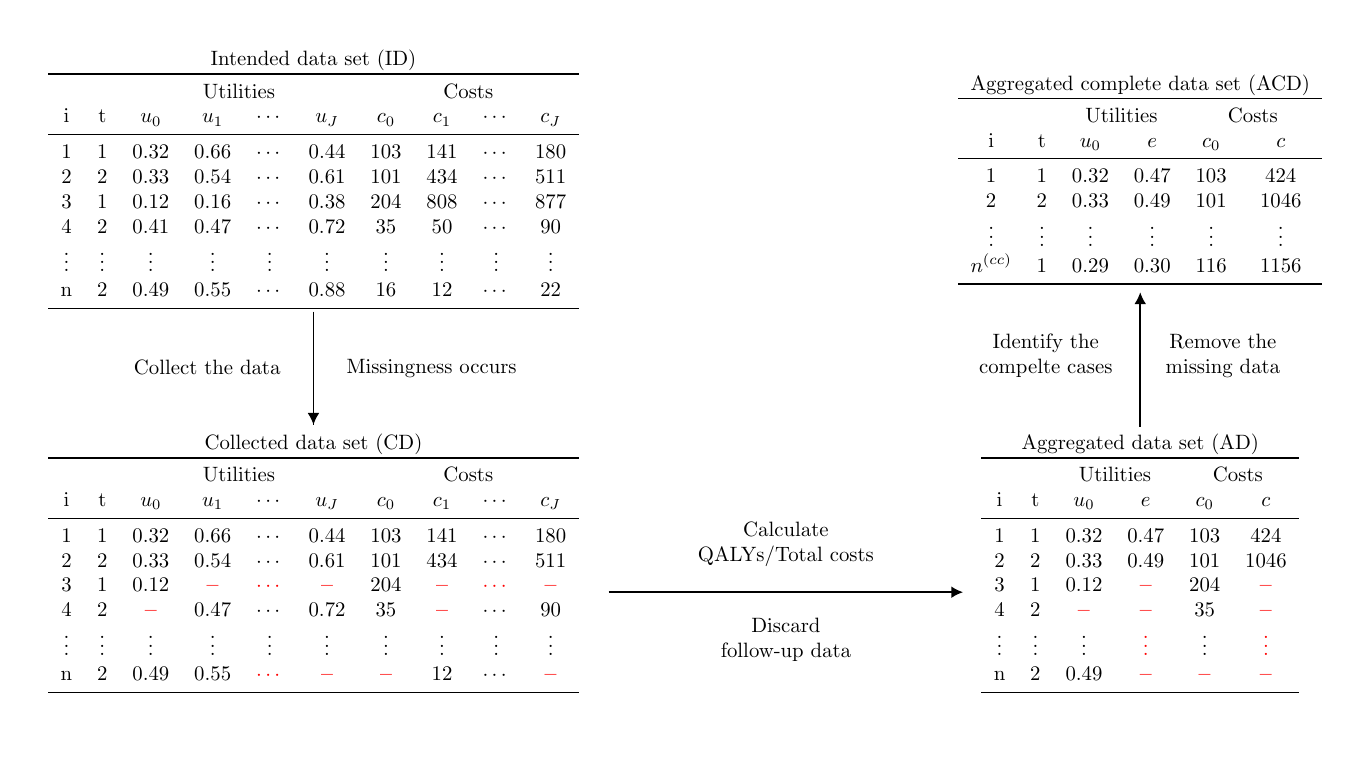}
\caption{\footnotesize Schematic representation of the standard procedure for processing trial-based CEA data. Starting from an ideal dataset, where all the variables that were intended to be collected are available (ID), in reality, some individuals drop out or do not show up at some scheduled time point, leading to a partially-observed collected dataset (CD). Next, aggregated quantities (e.g.~QALYs and total costs) are calculated for only those individuals with fully-observed utility and cost data, and the follow-up values are discarded to obtain an aggregated dataset (AD). Finally, all individuals with some missing values are removed in the aggregated complete dataset (ACD), which only includes individuals with fully-observed data.}
\label{structure}
\end{figure}

The standard practice in trial-based CEAs is to conduct the analysis (and apply the missingness method) at the level of the aggregated outcomes (e.g.~QALYs/total costs) and baseline variables (e.g.~baseline utilities/costs). This is due to the fact that the estimates of interest can be obtained by directly modelling the aggregated outcomes rather than modelling the utility and cost data at each time point in the study. This requires the analyst to process the utility and cost data collected on individual $i$ at time point $j$ in treatment group $t$, in order to derive the aggregated outcomes over the study duration. Figure~\ref{structure} shows a typical dataset of trial-based CEA, formed by the sets of utility $u_j$ and cost $c_j$ variables collected at baseline $j=0$ and some follow-ups $j=1,\ldots,J$. The graph provides a schematic representation of the standard procedure for processing the data and identify the variables to be used in the CEA. 
At the top-left of the diagram, the \textit{intended dataset} (ID) is displayed, which represents the ideal scenario where all the data are collected as originally intended for all the individuals in the trial. In real cases, however, it is likely that some individuals drop out from the study or are associated with some unobserved utility and cost values at some time point, which leads to the \textit{collected dataset} (CD). At this point, aggregated quantities for both the effectiveness ($e$) and cost ($c$) outcomes are computed, e.g.~using the area under the curve method to calculate the QALYs~\citep{Drummond}. These aggregated quantities can be calculated only for those subjects with fully-observed utility and cost data in the trial. The focus of the analysis is then moved from the original longitudinal dataset to the \textit{aggregated dataset} (AD), where all follow-up data are discarded and the only variables kept for the analysis are the baseline values and the aggregated outcomes. Finally, the standard approach in many cases is to remove all individuals with some missing values in either the baseline or aggregated variables and restrict the analysis to the \textit{aggregated complete dataset} (ACD), formed by the $n^{(cc)}<n$ individuals with fully-observed~data.

\subsection{Joint Longitudinal Models to Deal with Missingness}\label{jlm}
While a considerable amount of methodological work for trial-based CEAs has been carried out to handle missingness at an aggregated level (i.e.~based on the AD), particularly with the use of joint models~\citep{Mason2018,Gabriob,gomes2019copula}, the consideration of the missingness problem from a longitudinal perspective (i.e.~based on the CD) has been less extensively addressed. Only recently, \citet{gabrio2019bayesian} proposed a Bayesian longitudinal model to assess the impact of alternative missingness assumptions for trial-based CEAs. A general limitation of any method (including joint models) which addresses missingness in the aggregated outcomes is to ignore the longitudinal nature of the data and discard all partially-observed follow-up values. Conversely, the extension of these models to handle missingness in the utility and cost variables collected at each time point allows to properly account for the longitudinal nature of the data, incorporate all available evidence into the analysis, and thus potentially make the missingness assumptions (e.g.~MAR) more reasonable to justify. 

In this study, we illustrate the impact that alternative missing data approaches in trial-based CEAs have on the final inferences and cost-effectiveness conclusions. In particular, focus is given to how \textit{joint longitudinal models} can be used to extend aggregated models for dealing with the longitudinal nature of trial-based CEA data under MAR, using either multiple imputation (L-MI) or Bayesian methods (L-FB). This approach improves the current practice by taking into account the information from all observed utility and cost data as well as the time dependence between the responses. The economic analysis is then performed on the targeted quantities (i.e.~mean QALYs and total costs), which are retrieved by combining the mean utility and cost estimates obtained from the model. We compare the results obtained from alternative missing data methods using a simulation study and two real case studies. Inferences are obtained under MAR as we want to assess the impact on cost-effectiveness conclusions under the default missingness assumption in routine analyses.
The rest of the paper is structured as follows. Section~\ref{methods} briefly outlines the different missing data approaches considered, with a focus on how joint aggregate models can be extended to deal with the longitudinal framework of CEA studies. Section~\ref{models} summarises the assumptions and the implementation details for each approach used. The design and results from a simulation study are presented in Section~\ref{simulation}, while Section~\ref{results} shows the statistical and economic results from the two case studies. Finally, Section~\ref{discussion} concludes with a discussion.

\section{Methods}\label{methods}
This section briefly revises some of the most popular missing data methods in trial-based CEAs, within the standard aggregated modelling framework, which include: case deletion (CCA and ACA), mean baseline imputation (MEAN) and joint aggregated (MI and FB) methods. Joint longitudinal methods (L-MI and L-FB) are then introduced and discussed. Throughout the paper, to ease notation, we describe the methods under the simplified scenario where only two treatment groups are compared (e.g.~a new intervention and a control) and the only covariates are the baseline utilities and costs. However, we note that the methods can be easily extended to deal with multiple treatments and different types of baseline covariates, such as prognostic factors or demographic~variables.  

\subsection{Case Deletion and Baseline Imputation Methods}\label{cd_bi}
The standard approach in trial-based CEA fits the model of interest to the ACD or AD using the individual-level aggregated effectiveness and cost variables ($e_i, c_i$), which are computed after processing the utility and cost data collected from each time point in the study ($u_{ij}, c_{ij}$). In many cases, linear regression models are used to obtain treatment specific estimates and to adjust for baseline values, often under simplifying assumptions such as normality and independence between the two types of outcomes:
\begin{nalign}\label{datamodel}
e_{i} \mid t_i,u_{i0}&\sim \text{Normal}\left(\alpha_{0}+\alpha_{1}t_{i}+\alpha_{2}u_{i0}, \sigma^2_e \right)\\
c_{i} \mid t_i,c_{i0}& \sim \text{Normal}\left(\beta_{0}+\beta_{1}t_{i}+\beta_{2}c_{i0}, \sigma^2_c\right),
\end{nalign}
where $t_{i}$ is a treatment indicator variable, $\bm \alpha=(\alpha_0,\alpha_1)$ and $\bm \beta=(\beta_0,\beta_1)$ are the sets of regression parameters, and $\bm \sigma=(\sigma_e,\sigma_c)$ are the standard deviations indexing the effectiveness and cost models. Once the models are fitted and parameter estimates are derived, e.g.~using maximum likelihood or Bayesian methods, treatment specific population means $(\mu_{et},\mu_{ct})$ can be obtained by plugging-in the parameter estimates for $\bm \alpha$ and $\bm \beta$ into the linear formulae in Equation~\ref{datamodel} and replacing the baseline variables with their empirical means. When confronted with missing values in either or both the aggregated and baseline variables, different parameter estimates can be obtained according to the approach used to deal with missingness. For example, both CCA and ACA fit the models only to the complete cases and generate the mean estimates using the mean of the baseline variables computed from the complete or all observed cases, respectively. Baseline imputation methods, such as MEAN, fit the models to the entire dataset using single-imputed baseline variables ($u^\star_{i0},c^\star_{i0}$) and predict the missing values in $e_i$ and $c_i$ using the parameter estimates and the linear formulations, either ignoring (with point predictions) or accounting (by adding some random term) for the uncertainty around the missing outcome values. In all analyses based on MEAN, we impute the missing outcome values including the random term to properly take into account missing data uncertainty. 

\subsection{Joint Aggregated Models}\label{joint}
Joint aggregated models simultaneously handle missingness in both the aggregated and baseline variables through the specification of a joint (often Normal) distribution for $(e_i,u_{i0})$ and $(c_i,c_{i0})$. Sometimes it is more convenient to represent these joint models using conditional probabilities and factor the two bivariate distributions into the product of marginal and conditional distributions~\citep{Nixon}. This factorisation allows to re-express the joint models using univariate regressions, that is:
\begin{nalign}\label{datamodel_agg}
u_{i0} & \sim \text{Normal}\left(\mu_{u_{0}},\sigma^2_{u_{0}}\right), & e_{i} \mid t_i,u_{i0} \sim \text{Normal}\left(\alpha_{0}+\alpha_{1}t_{i}+\alpha_{2}u_{i0}, \sigma^2_{e}\right)\\
c_{i0} & \sim \text{Normal}\left(\mu_{c_{0}},\sigma^2_{c_{0}}\right), & c_{i} \mid t_i,c_{i0} \sim \text{Normal}\left(\beta_{0}+\beta_{1}t_{i}+\beta_{2}c_{i0}, \sigma^2_{c}\right),
\end{nalign}
where $(\mu_{u_{0}},\sigma_{u_{0}})$ and $(\mu_{c_{0}},\sigma_{c_{0}})$ denote the population means and standard deviations of the baseline utilities and costs, respectively. In Equation~\ref{datamodel_agg}, the coefficients $\alpha_2=\frac{\sigma_e}{\sigma_{u_{0}}}\rho_e$ and $\beta_2=\frac{\sigma_c}{\sigma_{c_{0}}}\rho_c$ quantify the association between the aggregated and baseline variables for the two outcomes, where $\rho_e$ and $\rho_c$ are correlation parameters. Different approaches, such as MI and FB~\citep{Little1992}, can be used to fit these joint models and retrieve the mean population estimates for the aggregated effectiveness and cost outcomes as described in Section~\ref{cd_bi}. 

\subsection{Joint Longitudinal Models}\label{long}
Joint aggregated models can be extended in a relatively easy way to account for the longitudinal nature of the data and use all the available utility and cost values in the CD. Assumptions about the longitudinal structure of the data can be made to simplify the specification and implementation of the models, either using MI or FB. For example, one of the simplest specification assumes a first-order Markov dependence for the utility and cost models, that is:
\begin{nalign}\label{datamodel_long}
u_{i0} & \sim \text{Normal}\left(\mu_{u_{0}},\sigma^2_{u_{0}}\right), & u_{ij} \mid t_i,u_{ij-1} \sim \text{Normal}\left(\alpha_{0jt}+\alpha_{1jt}u_{ij-1}, \sigma^2_{u_j}\right)\\
c_{i0} & \sim \text{Normal}\left(\mu_{c_{0}},\sigma^2_{c_{0}}\right), & c_{ij} \mid t_i,c_{ij-1} \sim \text{Normal}\left(\beta_{0jt}+\beta_{1jt}c_{ij-1}, \sigma^2_{c_j}\right),
\end{nalign}
where $({\alpha}_{jt},\sigma_{u_j})$ and $({\beta}_{jt},\sigma_{c_j})$ are the treatment-specific and standard deviation parameters indexing the models for the utility and cost variables at time $j$, conditional on their values at time $j-1$, respectively. Alternative specifications of the models are possible and, in theory, a multivariate model can be specified to account for the potential dependence across all variables and time points. However, we believe that the proposed specification offers a reasonable compromise which allows to capture key dependences between the utility and cost variables, while also providing a model that is relatively easy to implement in practice. 

When the models in Equation~\ref{datamodel_long} are fitted using MI, the joint distributions can be approximated by separately specifying the imputation and analysis steps. We denote this approach as L-MI, where the prefix "L" is used to indicate that the model is fitted in a longitudinal setting. First, a set of $M$ imputed values for $u_{i0}$ and $c_{i0}$ are generated; next the utilities and costs at at time $j>0$ are imputed including the utilities and costs at time $j-1$ in the corresponding imputation models. Finally, the models are fitted to each imputed dataset and the parameter estimates are pooled across the datasets. If a FB approach is used instead, the models in Equation~\ref{datamodel_long} can be directly fitted to the partially-observed utilities and costs, with imputed values that can be iteratively generated via data augmentation using MCMC algorithms. We denote this approach as L-FB, for which estimates based on the observed data (i.e.~MAR) can be obtained through the specification of weakly informative priors on all parameters. 

Regardless of whether the models are fitted using L-MI or L-FB, estimates of the mean aggregated effectiveness and costs in each treatment group can be retrieved in two steps. First, the estimates of the marginal means of the utilities and costs for each time and treatment group ($\mu_{ujt}$ and $\mu_{cjt}$) can be derived from Equation~\ref{datamodel_long}, for example by setting the utility and cost variables in the regressions to their sample mean values for each treatment group. Second, the standard formulae used for the calculation of $e_i$ and $c_i$ can be applied to $\mu_{ujt}$ and $\mu_{cjt}$ to directly obtain the estimates of the aggregated means $\mu_{et}$ and~$\mu_{ct}$. 

\section{Implementation}\label{models}

\subsection{Model Specification}
Table~\ref{table_models} summarises the key features and assumptions for each missingness approach described in Section~\ref{methods}.
\begin{table}[!h]
\centering
\begin{tabular}{r|cccc}
\toprule
method & modelled variables & dataset & imputation\\
\midrule
 \multirow{2}{*}{CCA} &  $e_i\mid t_i,u_{i0}$  &  \multirow{2}{*}{ACD} & \multirow{2}{*}{none} \\
 & $c_i\mid t_i,c_{i0}$ &   & \\[0.7em]
 \multirow{2}{*}{ACA} &  $e_i\mid t_i,u_{i0} $  &  \multirow{2}{*}{ACD} & \multirow{2}{*}{none} \\
 & $c_i\mid t_i,c_{i0}$ &   & \\[0.7em]
 \multirow{2}{*}{MEAN} &  $e_i\mid t_i,u^\star_{i0}$  &  \multirow{2}{*}{AD} & mean for $u_{i0},c_{i0}$ \\
 & $c_i\mid t_i,c^\star_{i0}$ &   & regression for $e_i,c_i$\\[0.7em]
 \multirow{2}{*}{FB/MI} & $(u_{i0},e_i) \mid t_i$  &  \multirow{2}{*}{AD} & FB/MI for $u_{i0},e_i$ \\
 & $(c_{i0},c_i) \mid t_i$&   & FB/MI for $c_{i0},c_i$\\[0.7em]
 \multirow{2}{*}{L-FB/L-MI} & $(u_{i0},\ldots,u_{iJ}) \mid t_i$  &  \multirow{2}{*}{CD} & FB/MI for $u_{i0},\ldots,u_{iJ}$ \\
 & $(c_{i0},\ldots,c_{iJ}) \mid t_i$ &   & FB/MI for $c_{i0},\ldots,c_{iJ}$\\
\bottomrule
\end{tabular}
\caption{List of the different methods for handling missing data in trial-based CEAs which are compared. A total of five approaches are considered: complete (CCA) and available (ACA) case analysis, mean baseline imputation (MEAN), joint aggregated (FB/MI) and longitudinal (L-FB/L-MI) models. The methods are fitted to different types of datasets (ACD, AD or CD) and imputed variables (none, aggregated and baseline, or all collected variables.}\label{table_models}
\end{table}
With the exception of L-FB/L-MI, which are fitted to and generate imputations based on all partially-observed variables in the CD, all other methods are fitted only to the aggregated and baseline variables, either using the completers in the ACD (CCA and ACA) or all cases in the AD through some imputations of the missing data (MEAN and MI/FB). In the following sections, we assess the impact on the inferences associated with the implementation of these methods, first using a small simulation study (Section~\ref{simulation}) and then fitting the models to real data from two randomised trials (Section~\ref{results}). In the simulation study, to ensure the comparison of the results across the different approaches, we fit all models using a Bayesian approach and assess the performance of the methods in terms of bias and efficiency using the posterior distributions for the targeted quantities of the analysis. In the application to the case studies, we additionally fit the joint aggregated and longitudinal models using MI, and assess the impact on the inferences for the different approaches in terms of both parameter estimates and cost-effectiveness conclusions for the treatments under~consideration. In both studies, we perform the analysis under MAR, as this is the default assumption in many applications, and we focus on the comparison between joint aggregated models (the currently recommended approach in the literature) and joint longitudinal models, which represent our proposed strategy to improve current practice. We also include other methods for comparison and to assess the potential gains of using joint models for handling missingness with respect to simpler approaches which are popular among practitioners. 

In all analyses, Bayesian models are fitted using weakly informative prior distributions for each parameter to minimise the impact of undesired prior information on the results. Specifically, we choose Normal distributions centred at $0$ with a standard deviation of $1000$ for all the regression coefficients ($\bm \alpha$ and $\bm \beta$) and Uniform distributions between ($0,1000$) for standard deviation parameters ($\bm \sigma$). Prior sensitivity to alternative specifications for all parameters was conducted by varying the hyperprior values over the set $\{1000,10000,100000\}$ for the standard deviation of the normally distributed $\bm \alpha$ and $\bm \beta$ and the upper bound for the uniformly distributed $\bm \sigma$. Overall, results from all analyses were robust to these variations. MI models are fitted using MICE, setting the number of imputations to $M=20$, and generating bootstrap samples to replicate the sampling distribution of the targeted estimates across the imputed datasets. We assess the impact on the inferences of alternative methods using point and interval estimates around the targeted quantities of the analysis, i.e.~the mean aggregated effectiveness and total cost differences between treatment groups. More specifically, for all models fitted under a Bayesian framework (CCA, ACA, MEAN, FB and L-FB), $95\%$ credible intervals are calculated from the $2.5$ and $97.5$ percentiles of the posterior distributions of the parameters. For all models fitted with MI (MI and L-MI), confidence intervals are evaluated using the percentile-t bootstrap method, which improves the accuracy of the intervals calculated from bootstrapped replications in small samples~\citep{Efron,Brand}. 

\subsection{Software}\label{software}
All Bayesian models are implemented using \texttt{JAGS}~\citep{Plummer}, a program dedicated for Bayesian analysis using MCMC simulation. Specifically, we interface \texttt{JAGS} with the freely available statistical software \texttt{R}, using the package \texttt{R2jags}~\citep{Su}. We ran two chains with $20,000$ iterations per chain, using a burn-in of $10,000$, for a total sample of $20,000$ iterations for posterior~inference. Samples from the posterior distribution of the parameters of interest are then saved to the \texttt{R} workspace and used for producing relevant statistics and plots. MI models are fitted using MICE through the \texttt{R} the package \texttt{mice}~\citep{micepkg}. We first created $M=20$ imputed datasets and for each of these we generated $B=1000$ bootstrap samples, which allowed to obtain a total of $20,000$ values for each estimate of interest. For all models, the convergence of the algorithms was assessed using different types of diagnostic measures, such as the \textit{potential scale reduction factor}~\citep{Gelman2} or visual inspection of the trace plots of the model parameters (Bayesian) or imputed values (MI). As an example of how to implement each type of approach shown in Table~\ref{table_models}, the \texttt{JAGS} code used to fit all Bayesian models to the MenSS data is provided in the online supplementary material.

\section{Simulation Study}\label{simulation}
In this section we present the design and results of a simulation study whose aim is to provide a general picture on the performance of the missingness approaches described in Section~\ref{models} for different assumptions about the data generating and missingness process. We consider a simple randomised trial  setting with two treatments begin compared and only one type of outcome (the utilities), which is assumed to be normally distributed and collected at three different time points evenly spaced over one year (at baseline, $6$ and $12$ months).  

\subsection{Data Generating Process}\label{dataprocess}
Consider a two-arm randomised trial where a total of $n$ individuals are enrolled and for whom utility scores $u_j$ are available at baseline ($j=0$), $6$ months ($j=1$) and $12$ month follow-ups ($J=2$). We assume that the number of individuals assigned to each treatment group $t=(1,2)$ is the same (equal to $n/2$), and we omit the individual index $i$ in the following to ease notation. The data generating process for the utilities is specified through a multivariate normal distribution
\begin{nalign}\label{joint_N}
\bm u \sim  \text{Normal} 
\left(\begin{array}{c} 
\bm \mu_{u}\\
\end{array},
\begin{array}{ccc}  
\bm \Sigma_{u}
\end{array}\right),
\end{nalign}
where $\bm u=(u_0,u_1,u_2)$ is the set of the utility variables, while $\bm \mu_{u}$ and $\bm \Sigma_{u}$ are the mean vector and the $3\times3$ covariance matrix indexing the joint distribution of the utilities. We specify Equation~\ref{joint_N} by assuming treatment specific post-baseline mean parameters and a shared baseline mean and covariance matrix between the two treatment groups. Starting from a population mean of $\mu_{u_{0}}=0.4$ for all participants, we assume that the individuals in the first treatment $(t=1)$ show an average utility improvement which remains constant between the follow-up points, with $\mu_{u_{1}}=\mu_{u_{2}}=0.5$. In the second treatment ($t=2$), the individuals start from the same mean baseline value but show a larger average utility improvement at $j=1$ which is further increased at $j=2$, with $\mu_{u_{1}}=0.6$ and $\mu_{u_{2}}=0.7$. In both treatments, the elements of $\bm \Sigma_{u}$ are set so that all individuals are associated with a constant population standard deviation $\sigma=0.1$ and correlation $0.5$ for the utilities across the time points. We calibrated the values of the parameters to obtain simulated CEA data that mimic to the structure of the data observed in one of the two cases studies used as motivating examples (Section~\ref{pbs_t}). This data generating process leads to mean aggregate QALYs of $\mu_{e1}=0.475$ and $\mu_{e2}=0.575$ over the trial duration in the first and second intervention group, respectively (i.e.~corresponding to a treatment effect of $\Delta=0.1$~QALYs).

\subsection{Missingness Mechanism}\label{missprocess}
We model the missingness mechanism for the utilities under the assumption that missingness is due to dropout, i.e.~only monotone patterns, and that is the same between the two treatment groups. Specifically, individuals are assumed to either complete the study with no missing values (completers) or to drop out at~$j \in \{0,1,2\}$. At each time point, we specify the model for the missing data indicators $m_{u_{j}}$ (taking value $1$ if $u_j$ is missing and $0$ otherwise) using Bernoulli distributions and logit link functions to model the linear dependence between the probability of missingness and the utility variables based on different types of mechanisms. At baseline, missingness is generated~as
\begin{nalign}\label{mmu0}
m_{u_{0}}&\sim \text{Bernoulli}(\pi_{u_{0}})\\
\text{logit}(\pi_{u_{0}})&= \eta_{00}+ \eta_{10}u_{0},
\end{nalign}
where $\pi_{u_{0}}$ is the missingness probability at $j=0$, while $\bm \eta_0=(\eta_{00},\eta_{10})$ is the set of parameters indexing the linear logit regression. Since missingness is assumed to be monotone, at the first follow-up, for each individual with $m_{u_{0}}=1$, the value of $u_{1}$ is set missing. Among the subset of the individuals still in the study at $j=1$, missingness is generated as
\begin{nalign}\label{mmu1}
m_{u_{1}}&\sim \text{Bernoulli}(\pi_{u_{1}})\\
\text{logit}(\pi_{u_{1}})&= \eta_{01}+ \eta_{11}u_{0}+\eta_{21}u_{1},
\end{nalign}
where $\pi_{u_{1}}$ is the missingness probability at $j=1$, while $\bm \eta_1=(\eta_{01},\eta_{11},\eta_{21})$ is the set of the regression parameters. Finally, at the second follow-up, among the subset of the individuals who did not drop out at $j<2$, missingness is generated as 
\begin{nalign}\label{mmu2}
m_{u_{2}}&\sim \text{Bernoulli}(\pi_{u_{2}})\\
\text{logit}(\pi_{u_{2}})&= \eta_{02}+ \eta_{12}u_{0}+\eta_{22}u_{1}+\eta_{32}u_{2},
\end{nalign}
where $\pi_{u_{2}}$ is the missingness probability at $j=2$, while $\bm \eta_2=(\eta_{02},\eta_{12},\eta_{22},\eta_{32})$ is the set of regression~parameters. 

\subsection{Design of the Simulation}\label{design}
We design the simulation study using different specifications for the data generating process and missingness mechanism to compare the performance of the missing data approaches described in Section~\ref{models} over a range of alternative scenarios. More specifically, each scenario is defined by varying two types of parameters: the sample size ($n$) of each simulated dataset and the values of the parameters indexing the missingness models ($\bm \eta_0$, $\bm \eta_1$ $\bm \eta_2$). The latter are varied over a discrete grid to obtain different configurations in terms of the proportions of missing values and type of missing data mechanism. We consider a total of $45$ simulation scenarios, which are defined based on the combination of:
\begin{itemize}
\item Three values of sample size: $100$, $500$ and $1000$.
\item Three values of average proportion of missing utilities over the study period: a "low" rate of $0.15$, a "medium" rate of $0.30$, and a "high" rate of $0.5$.
\item Five types of missingness mechanisms: MCAR, and two alternative definitions of MAR and~MNAR.
\end{itemize} 
Table~\ref{tabdesign} shows the values for all non-zero parameters indexing the models of the missing data indicators, which are associated with the five missingness mechanisms considered (reported for the scenario with the "medium" proportion of missing values). The mechanisms for the scenarios with "low" and "high" missing data proportions are obtained by simply calibrating the values of the intercept parameters to match the desired proportions.
\begin{table}[!h]
\centering
\scalebox{1}{%
\begin{tabular}{r|c}
  \toprule
 \multicolumn{2}{c}{\textbf{Missingness Mechanism}}  \\ 
  \midrule
 mechanism & parameters  values \\ 
\midrule
MCAR & $\eta_{00}=-2$, $\eta_{01}=-2$, $\eta_{02}=-2$  \\ 
\midrule
 \multirow{2}{*}{MAR1} & $\eta_{00}=-2$, $\eta_{01}=-5$, $\eta_{02}=-5$,  \\ 
  & $\eta_{11}=8,\eta_{12}=8$ \\
\midrule
 \multirow{2}{*}{MAR2} & $\eta_{00}=-2$, $\eta_{01}=-6$, $\eta_{02}=-6$,  \\ 
  & $\eta_{11}=8,\eta_{22}=8$ \\
\midrule
 \multirow{2}{*}{MNAR1} & $\eta_{00}=-2$, $\eta_{01}=-7.5$, $\eta_{02}=-7.5$, \\ 
  & $\eta_{10}=8,\eta_{21}=8,\eta_{32}=8$ \\
\midrule
 \multirow{2}{*}{MNAR2} & $\eta_{00}=-2$, $\eta_{01}=-7.3$, $\eta_{02}=-7.3$,  \\ 
  & $\eta_{21}=8, \eta_{32}=8$\\
   \bottomrule
\end{tabular}
}\caption{Values of the parameters indexing the models for the missing data indicators across five alternative missingness mechanisms, reported for the scenario with an overall "medium" proportion of missing utilities (equal to~$0.3$). For each mechanism, only the values of the parameters which are different from zero are~reported for clarity.}\label{tabdesign}
\end{table}
All mechanisms, with the exception of MCAR, assume that individuals associated with higher utility values are more likely to drop out from the study compared with those with lower values (as indicated by positive values of the non-intercept parameters). The mechanisms can be interpreted as follows. MCAR: drop out probabilities are totally random at any time $j$; MAR1 and MAR2: drop out at $j=0$ is totally random, while at $j=1,2$ it is more likely among individuals with higher values either at $j=0$ (MAR1) or at $j-1$ (MAR2); MNAR1 and MNAR2: drop out at $j=1,2$ is more likely among individuals with higher values at the same time, while drop out at $j=0$ is either totally random (MNAR1) or more likely for those with higher values at the same time (MNAR2). 

We design these mechanisms with the aim of assessing the performance of the methods under a framework which tries to mimic that from the two case studies analysed in Section~\ref{results}. In particular, in both studies the interventions assessed are non life-threatening and the target populations have a relatively good quality of life. Thus, if missingness is MNAR, we believe it is likely that individuals who drop out are those associated with better health states, who may decide to leave the study due to the limited impact the interventions have on their quality of life, with respect to those who remain in the study. However, we acknowledge that alternative scenarios could be investigated in a more comprehensive simulation study. For example, missingness could be defined such that individuals with either higher or lower utilities (or a combination of both) are more likely to drop out, or alternative forms for the missingness models could be considered (e.g.~using probit regressions or including interaction terms). However, since an extensive simulation is not the main objective of this paper, we decided to investigate a limited number of scenarios that, we believe, could be most relevant for replicating the framework of the case studies analysed.

For each scenario, we fit the models described in Table~\ref{table_models} to each simulated dataset using a Bayesian approach and repeat the process $S=500$ times. In each simulation, we derive and store the values of the mean QALYs differentials between the treatment groups estimated from each model, which are taken to be the mean evaluated over the posterior samples of the estimates. As part of the assessment of the models, we look at the bias and empirical standard errors of the estimates of interest in each scenario. We define the bias and empirical standard errors for $\Delta$~as:
\begin{equation}\label{assess}
\text{bias}=\frac{1}{S}\sum_{s=1}^S(\hat{\Delta}_{s} - \Delta) , \;\;\;  \;\;\; \text{empirical se}=\sqrt{\frac{1}{S-1}\sum_{s=1}^S(\hat{\Delta}_s - \bar{\Delta})^2},
\end{equation}
where $\hat{\Delta}_s$ is the estimate from one of the models at simulation $s$, $\bar{\Delta}$ is the average values of the estimates over the total number of simulations $S$, and $\Delta$ is the value of the parameter used in the~simulation. For both bias and empirical standard errors, point and $95\%$ interval estimates are computed based on the empirical distributions of the posterior samples of the parameters.

\subsection{Results of the Simulation}\label{results_sim}
\begin{figure}[!h]
\centering
\includegraphics[scale=0.6]{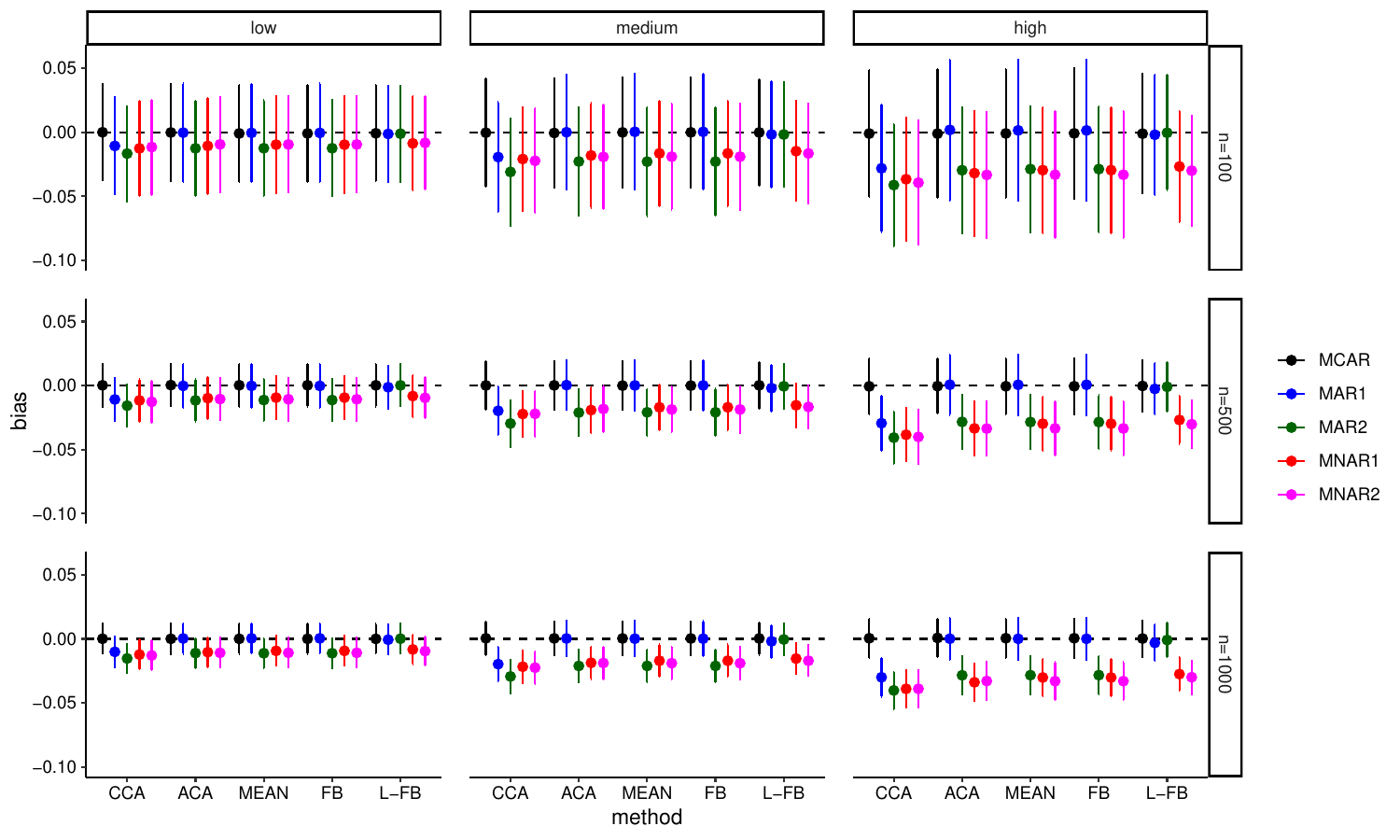}
\caption{Simulation study: bias associated with five missingness methods (CCA, ACA, MEAN, FB, L-FB) across $45$ scenarios defined according to: three values of missingness rate (low$=0.15$, medium$=0.3$ and high$=0.5$ - denoted by columns); three values of sample size ($n=100$, $n=500$ and $n=1000$ - denoted by rows); five types of missingness mechanisms (MCAR, MAR1, MAR2, MNAR1, MNAR2 - denoted with different colours). Results from each method are displayed in terms of average bias (points) and $95\%$ credible intervals (solid lines), computed over a total of $500$ simulations for each scenario. Absence of bias is indicated with black dotted lines.}\label{bias}
\end{figure}

\begin{figure}[!h]
\centering
\includegraphics[scale=0.6]{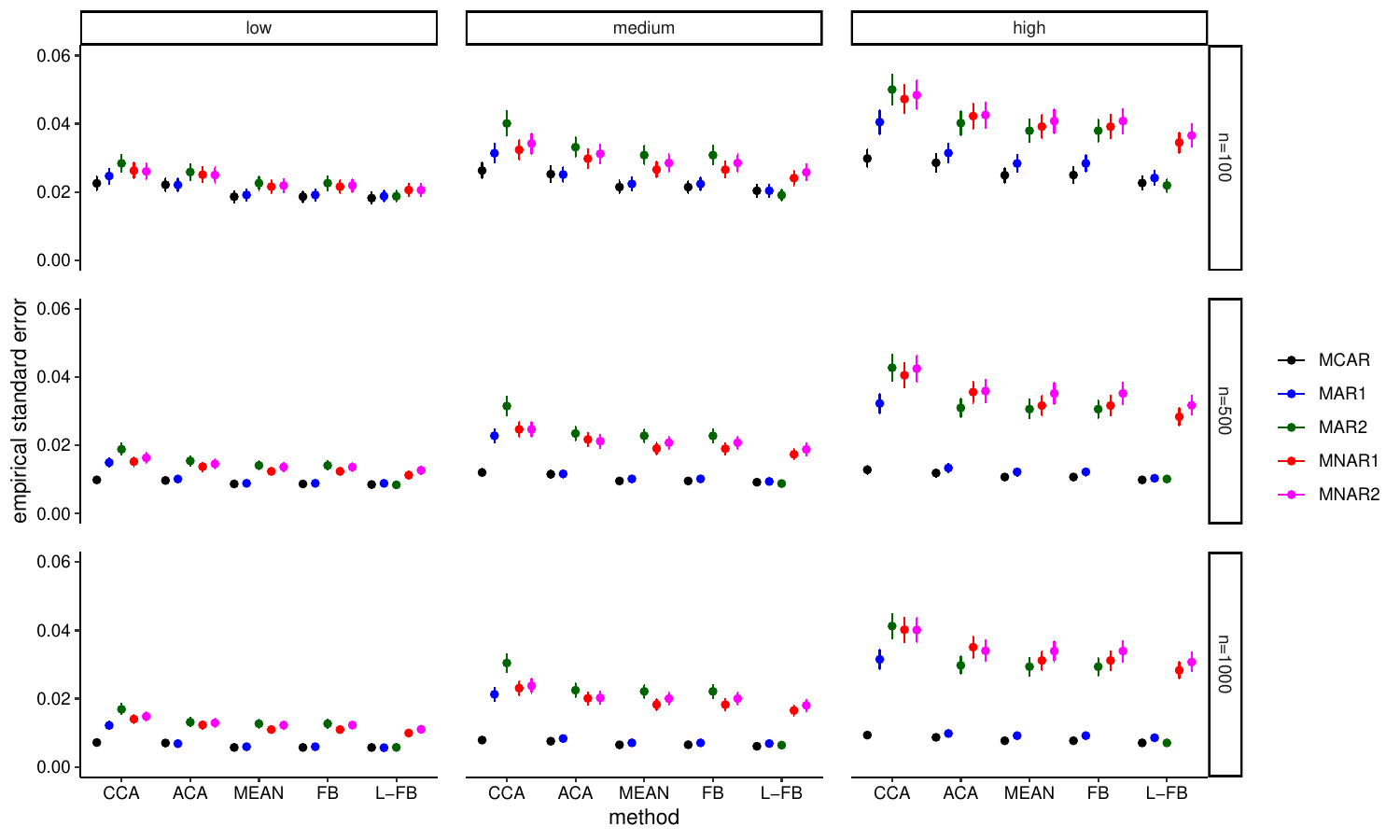}
\caption{Simulation study: empirical standard error associated with five missingness methods (CCA, ACA, MEAN, FB, L-FB) across $45$ scenarios defined according to: three values of missingness rate (low$=0.15$, medium$=0.3$ and high$=0.5$ - denoted by columns); three values of sample size ($n=100$, $n=500$ and $n=1000$ - denoted by rows); five types of missingness mechanisms (MCAR, MAR1, MAR2, MNAR1, MNAR2 - denoted with different colours). Results from each method are displayed in terms of average empirical standard error (points) and $95\%$ credible intervals (solid lines), computed over a total of $500$ simulations for each scenario. In those scenarios where intervals are particularly small with respect to those from other scenarios, only point estimates can be discerned.}\label{se}
\end{figure}

Results of the simulation study for each missingness method and scenario explored are summarised graphically in Figure~\ref{bias} and Figure~\ref{se} in terms of bias and empirical standard errors, respectively. As expected, under MCAR (black lines and dots), all methods show, on average, unbiased estimates and the smallest standard errors with respect to the other types of  mechanisms. Under MAR1 (blue lines and dots), where follow-up missingness depends on baseline data, CCA is the only method which shows biased estimates and higher standard errors compared to MCAR, while all other methods show comparable good performances. 
Under MAR2 (green lines and dots), where follow-up missingness at a given time depends on the observed data at the previous time, all methods that ignore the longitudinal nature of the data (CCA, ACA, MEAN and FB) show, on average, biased estimates and are associated with an increase in the standard errors, with CCA having the worst performance for both measures. L-FB is the only method showing unbiased estimates and standard errors which are comparable with those under MCAR and MAR1. Finally, under both MNAR mechanisms, where missingness depends on the unobserved values at the same time either including (MNAR1 - red lines and dots) or excluding (MNAR2 - magenta lines and dots) baseline data, all methods show, on average, biased estimates and higher standard errors with respect to those obtained for unbiased results. This is not surprising as all methods make inference based on the assumption that either missingness can be completely ignored (CCA, ACA) or that all necessary information is contained in the observed data at baseline (MEAN, FB) or at all times (L-FB). However, under MNAR mechanisms, this is not enough and the inclusion of external information about missingness (i.e.~not from the observed data) is needed to obtain correct inferences.
The results under each type of mechanism are in general comparable with few variations across the scenarios. The length of the intervals for both bias and empirical standard errors increases as the sample size decreases, while the magnitude of the bias and the standard errors increases as the proportion of missingness is increased.

\section{Application}\label{results}
In this section, for each of the two case studies used as motivating examples, we briefly describe the data and compare the statistical and economic results obtained from the implementation of the missingness methods shown in Table~\ref{models} to these datasets. 

\subsection{The MenSS Trial}\label{menss_t}
The Men's Safer Sex~\citep[MenSS;][]{Hunterb} randomised trial is a pilot study of a new digital intervention (the MenSS website) aimed at increasing condom use and reducing the incidence of sexually transmitted infections (STI) in young men. Individuals ($n=159$) enrolled in the study are men aged $16$ or over who report female sexual partners and recent unprotected sex or suspected acute STI. Participants were randomised to receive the MenSS website plus usual clinic care (reference intervention, $n_{2}=84$), or usual clinic care only (control, $n_{1}=75$). Sexual health related resource use was collected via participant responses to questionnaires at $3$, $6$ and $12$ months. Utility scores to calculate QALYs were collected at baseline and at the same time intervals as costs using the EQ-5D instrument. Table~\ref{sumMenss} shows the summary statistics for the utilities and costs at each time point in the trial. The number of subjects, means and standard deviations are separately reported between individuals with fully-observed data (completers) and those with partially-observed data (non-completers) and by treatment group.  
\begin{table}[!h]
\centering
\resizebox{0.85\textwidth}{!}{%
\begin{tabular}{r|c|cccc|ccc}
  \toprule
 & $\bm n$ & $\bm u_0$ & $\bm u_1$ & $\bm u_2$ & $\bm u_3$ & $\bm c_1$ & $\bm c_2$ & $\bm c_3$ \\ [0.2em]
\midrule
& & mean (sd) & mean (sd) & mean (sd) & mean (sd) & mean (sd) & mean (sd) & mean (sd)\\
\textbf{Control (t=1)} &  &  &  &  &  &  &  &  \\ 
completers & 27 & 0.92 (0.11) & 0.92 (0.14) & 0.88 (0.18) & 0.91 (0.14) & 160 (246) & 36 (80) & 12 (45) \\ 
   non-completers & 48 & 0.86 (0.18) & 0.88 (0.09) & 0.83 (0.28) & 0.78 (0.36) & 129 (214) & 120 (167) & 42 (80) \\ [0.5em]
  \textbf{Intervention (t=2)} &  &  &  &  &  &  &  &  \\ 
  completers & 19 & 0.83 (0.25) & 0.9 (0.16) & 0.9 (0.15) & 0.95 (0.08) & 132 (120) & 20 (56) & 33 (61) \\ 
    non-completers & 65 & 0.88 (0.22) & 0.91 (0.1) & 0.82 (0.28) & 0.86 (0.27) & 215 (189) & 6 (11) & 5 (8) \\ 
   \bottomrule
\end{tabular}
}\caption{Empirical means and standard deviations of the utilities and costs at time $j=0$ (baseline) and $j=1,2$ and $3$ (corresponding to $3$, $6$ and $12$ months follow-ups) in the MenSS trial. Summary statistics are separately reported between completers and non-completers and by treatment group. Baseline data are only available for the~utilities, and costs are expressed in $\pounds$.}\label{sumMenss}
\end{table}
The proportions of complete cases is relatively small in both treatment groups (about $36\%$ in the control and $23\%$ in the intervention). We note that, with few exceptions, the non-completers in the control are associated with lower mean utilities and higher costs compared with the completers, while no clear pattern between the two is observed in the intervention. The complete list of the missingness patterns and histograms of the empirical distributions of the QALYs and total costs (computed from aggregating the utility and cost variables) in the MenSS trial are provided in Table~\ref{tabpatmenss} and Figure~\ref{data_Menss} in Appendix~\ref{A1}.

\subsubsection{Results}\label{res_menss}
\begin{figure}[!h]
\centering
\subfloat[QALYs]{\includegraphics[scale=0.4]{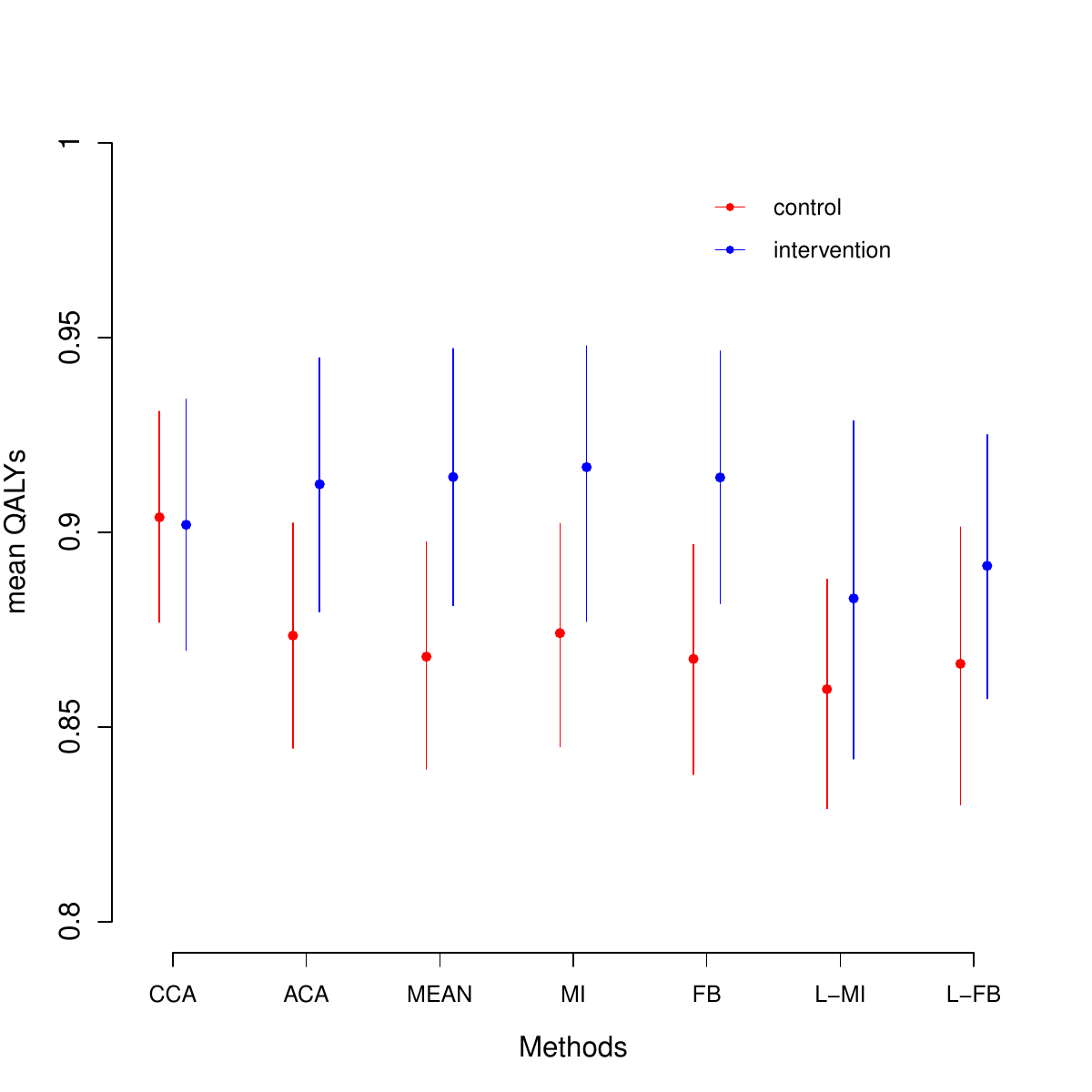}}
\subfloat[Total costs ($\pounds$)]{\includegraphics[scale=0.4]{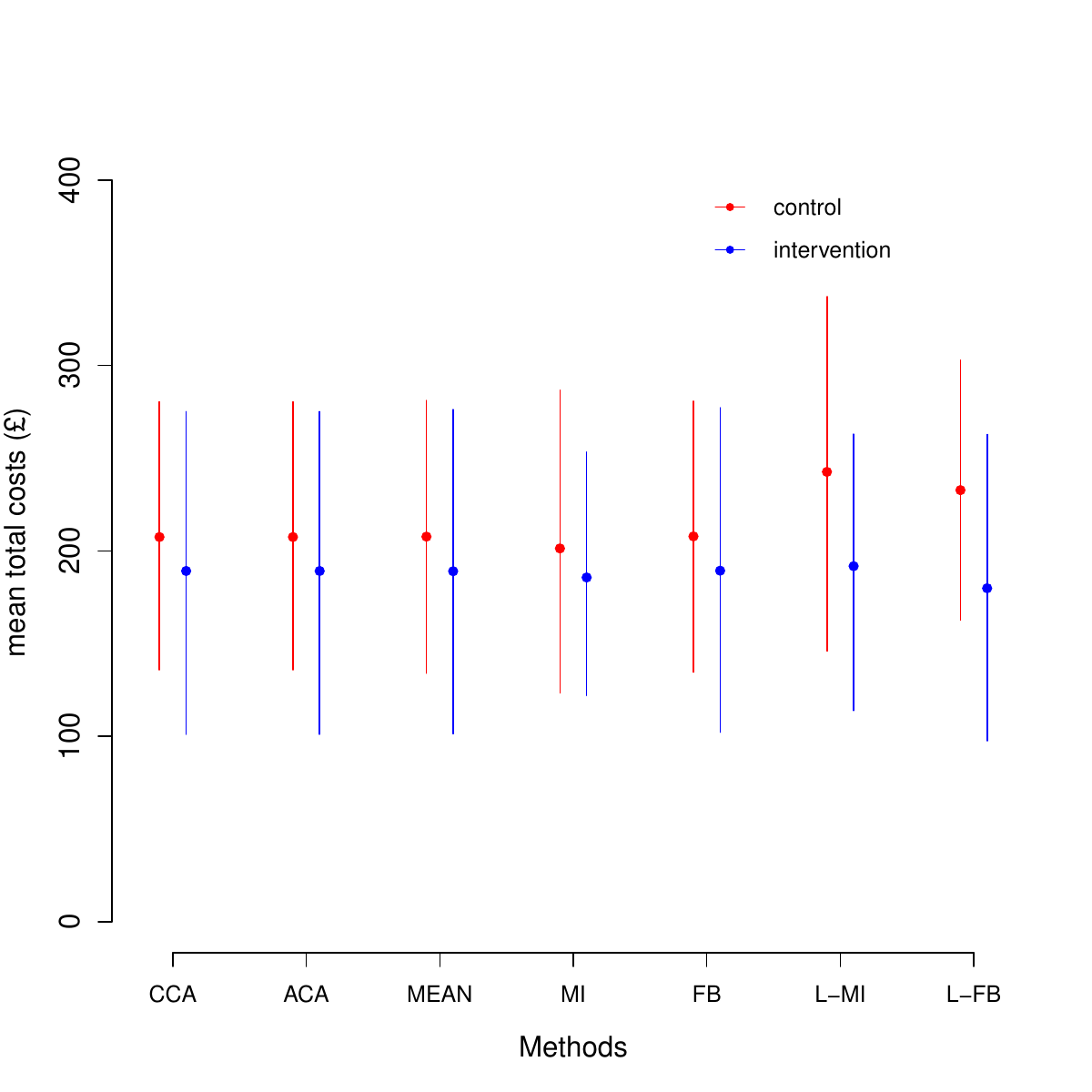}}
\caption{Mean and 95\% confidence/credible intervals for the mean QALYs (panel a) and total costs (panel b) in the MenSS trial, displayed by missingness method. The results for the control and intervention group are indicated with red and blue coloured dots and lines, respectively.}\label{fig_MENSS}
\end{figure}

Figure~\ref{fig_MENSS} displays the mean and 95\% confidence/credible intervals for the mean QALYs and total costs estimated from the application of seven different missingness methods, both in the control (red dots and lines) and intervention (blue dots and lines) group of the MenSS trial. We note that, since baseline costs were not collected in the study, for all methods, mean total costs estimates are derived without adjusting for these variables.
Considerable variations in the mean QALYs are observed across the methods. Under CCA, the mean QALYs difference between the two groups is near to zero. However, when the information from the observed values among the non-completers is incorporated into the model, either using ACA, MEAN or joint aggregated methods (MI and FB), the mean estimates in the intervention become systematically higher compared with those in the control. Similar conclusions are obtained from joint longitudinal models (L-MI and L-FB), even though mean QALY estimates under these approaches are shifted downwards in both treatment groups with respect to those from joint aggregated models. Mean total costs are similar across all methods fitted to the aggregated variables, while estimates from the joint longitudinal models are slightly higher, especially in the control group. For each method, we separately provide in Table~\ref{tabestmenss} in Appendix~\ref{app_menss} numerical summary statistics associated with different CEA quantities (including incremental results) for the analysis of the MenSS study. 

\begin{figure}[!h]
\centering
\textbf{Cost-Effectiveness Acceptability Curve}\par
\includegraphics[scale=0.45]{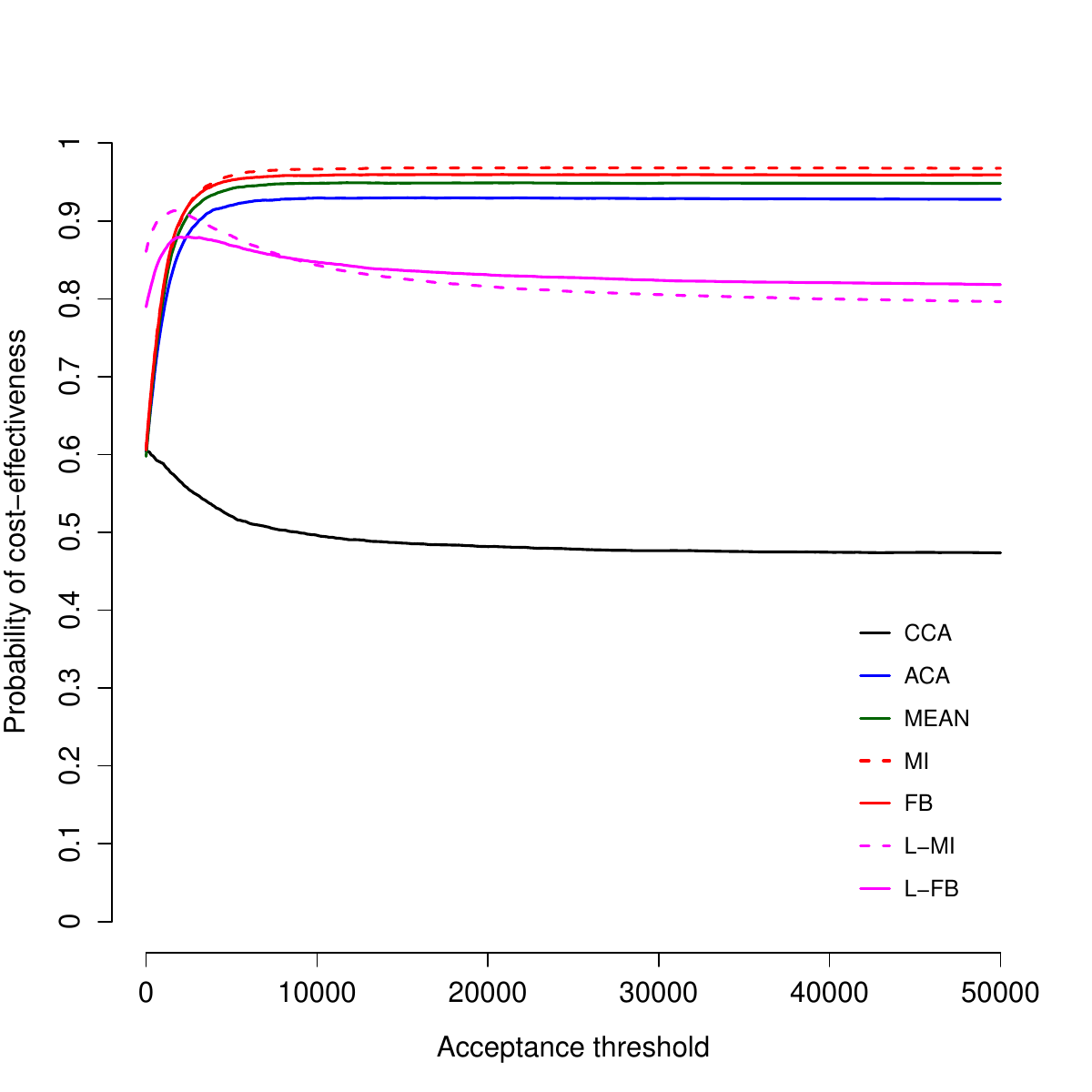}
\caption{Cost-effectiveness acceptability curves associated with seven alternative missingness approaches fitted to the MenSS study. Solid and dashed lines are used to indicate the results associated with the models fitted using Bayesian (CCA, ACA, MEAN, FB and L-FB) and multiple imputation methods (MI and L-MI). A variation of the acceptance thresholds up to \pounds{50,000} is considered.}\label{fig_MENSS_cea}
\end{figure}

We summarise the economic results from the trial by looking at the probability that the new intervention is cost-effective with respect to the control for different values of the acceptance threshold. Figure~\ref{fig_MENSS_cea} shows the cost-effectiveness acceptability curves~\citep[CEAC;][]{VanHout}, which are computed based on the posterior/bootstrapped samples of the mean QALYs and total costs, and which are associated with the methods applied to the MenSS study. The results are distinguished using different colours and types of lines: solid lines denote the methods fitted under a Bayesian framework (black for CCA, blue for ACA, green for MEAN, red for FB and magenta for L-FB), while dashed lines denote the methods fitted using multiple imputations (red for MI and magenta for L-MI). 
With the exception of CCA, all other methods indicate a relatively high probability of cost-effectiveness for the new intervention with respect to the control for most values of the acceptance threshold. In particular, all methods fitted to the aggregated variables, including joint models (MI and FB), suggest that the new intervention is cost-effective with a probability very close to one almost regardless of the value of the threshold. Joint longitudinal models (L-MI and L-FB), instead, indicate a milder conclusion with a slight decrease in the chance of cost-effectiveness down to $0.8$ at $\pounds 50,000$.

\subsection{The PBS trial}\label{pbs_t}
The Positive Behaviour Support \citep[PBS;][]{Hassiotis} is a multi-centre randomised trial involving community intellectual disability services and service users with mild to severe intellectual disability and challenging behaviour. PBS is a multicomponent intervention, which is designed to foster prosocial actions and enhance the person's quality of life and his/her integration within the local community. Participants ($n=244$) were enrolled from a total of $23$ sites and randomly allocated on a site basis to staff teams trained to deliver PBS in addition to treatment as usual (reference intervention, $n_{2}=108$), or to staff teams trained to deliver treatment as usual alone (control, $n_{1}=136$). Measures for quality of life (EQ-5D) and health related cost (family and paid carer records) were collected at baseline, $6$ and $12$ months. Table~\ref{sumPBS} reports the number of subjects, means and standard deviations for the completers and non-completers, separately for each treatment group in the trial.  
\begin{table}[!h]
\centering
\resizebox{0.85\textwidth}{!}{%
\begin{tabular}{r|c|ccc|cccc}
  \toprule
 & $\bm n$ & $\bm u_0$ & $\bm u_1$ & $\bm u_2$ & $\bm c_0$ & $\bm c_1$ & $\bm c_2$ \\ [0.2em]
\midrule
& & mean (sd) & mean (sd) & mean (sd) & mean (sd) & mean (sd) & mean (sd)\\
 \textbf{Control (t=1)} &  &  &  &  &  &  &  \\ 
  completers & 108 & 0.49 (0.37) & 0.5 (0.35) & 0.49 (0.33) & 1505 (2163) & 1448 (1846) & 1462 (3936) \\ 
  non-completers & 28 & 0.56 (0.33) & 0.51 (0.48) & 0.43 (0.33) & 2032 (2013) & 1606 (1379) & 679 (645) \\ [0.5em]
  \textbf{Intervention (t=2)} &  &  &  &  &  &  &  \\ 
   completers & 96 & 0.56 (0.39) & 0.64 (0.33) & 0.62 (0.32) & 2881 (1968) & 2845 (2210) & 2848 (1837) \\ 
   non-completers & 12 & 0.5 (0.29) & 0.68 (0.35) & 0.63 (0.42) & 3190 (1174) & -- (--) & 4781 (--) \\ 
   \bottomrule
\end{tabular}
}\caption{Empirical means and standard deviations of the utilities and costs at time $j=0$ (baseline) and $j=1$ and $2$ (corresponding to $6$ and $12$ months follow-ups) in the PBS trial. Summary statistics are reported for the completers and non-completers by treatment group. Costs are expressed in $\pounds$.}\label{sumPBS}
\end{table}
The proportions of complete cases is relatively large in both the control (about $80\%$) and intervention (about $90\%$) groups. Overall, no systematic differences are observed between the completers and non-completers, with the only exception given by the cost variables, especially in the intervention, where non-completers are associated with higher values. However, these differences are based on a quite small number of individuals ($28$ in the control and $12$ in the intervention) and should therefore be interpreted with care. The complete list of the missingness patterns and histograms of the empirical distributions of the QALYs and total costs in the PBS trial are provided in Table~\ref{tabpatpbs} and Figure~\ref{data_PBS} in Appendix~\ref{A1}.

\subsubsection{Results}\label{res_pbs}
\begin{figure}[!h]
\centering
\subfloat[QALYs]{\includegraphics[scale=0.45]{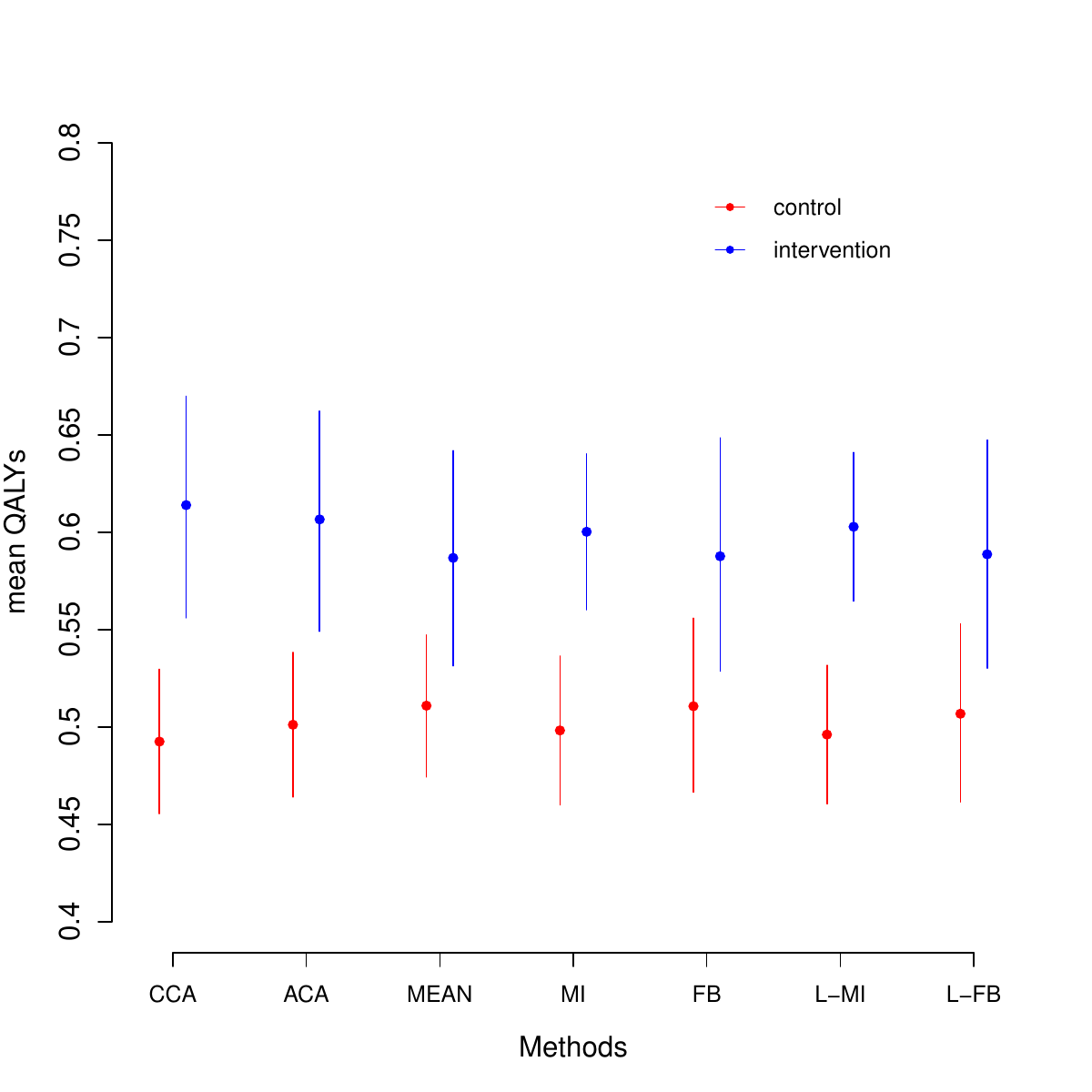}}
\subfloat[Total costs ($\pounds$)]{\includegraphics[scale=0.45]{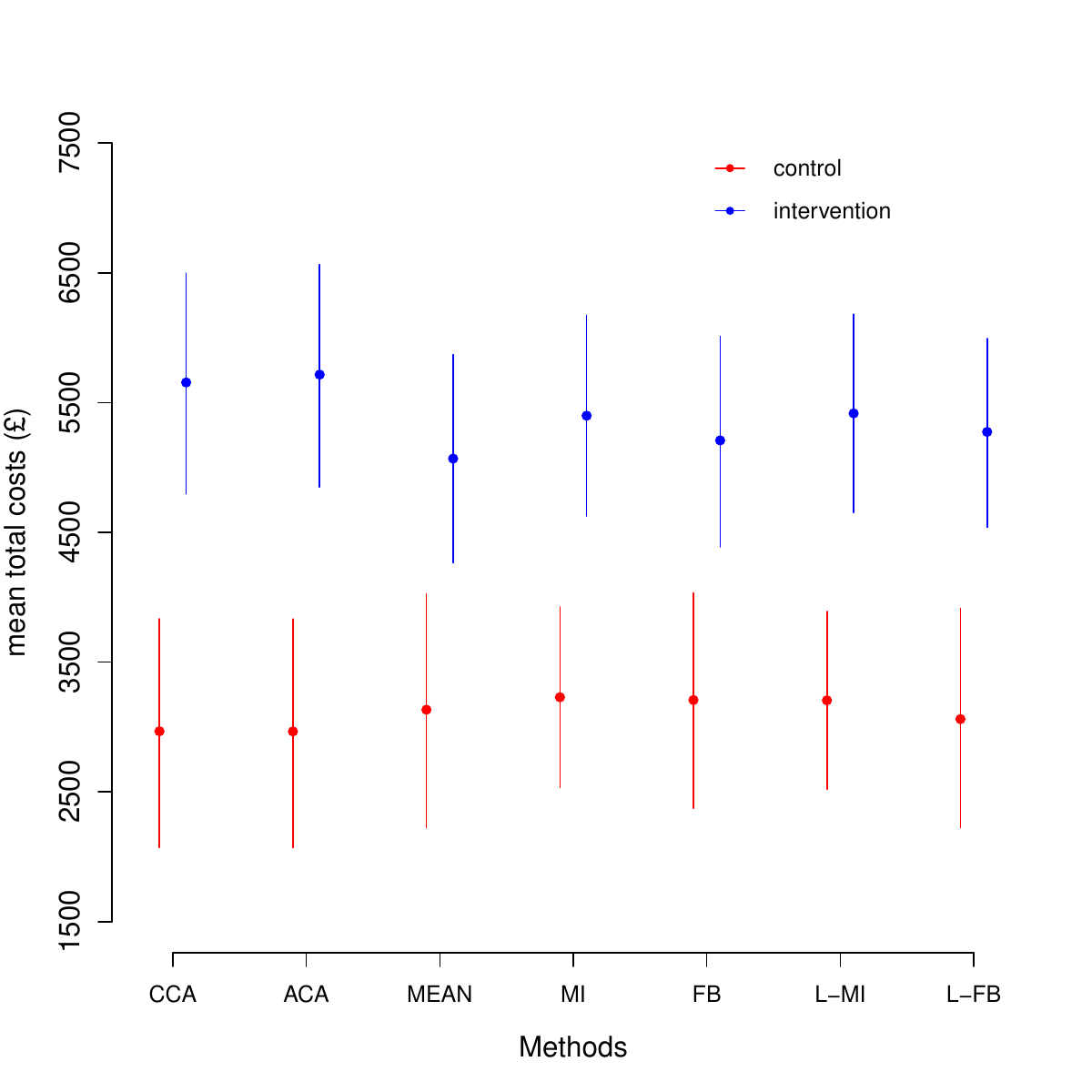}}
\caption{Mean and 95\% confidence/credible intervals for the mean QALYs (panel a) and total costs (panel b) in the PBS trial, displayed by missingness method. The results for the control and intervention group are indicated with red and blue coloured dots and lines, respectively.}\label{fig_PBS}
\end{figure}

Figure~\ref{fig_PBS} displays the mean and 95\% confidence/credible intervals for the mean QALYs and total costs estimated from each missingness method, both in the control (red dots and lines) and intervention (blue dots and lines) group of the PBS trial. For all methods, mean QALYs and total costs estimates are derived by taking into account the clustering in the data using random intercept models. 
Estimates for the mean QALYs and total costs do not largely vary across the methods, and indicate that the new intervention is both more effective and more expensive than the control. Results from CCA and ACA indicate the highest (lowest) mean estimates for the intervention (control) group among all methods, while relatively small differences are observed between the estimates from MEAN, the joint aggregated models (MI and FB) and the joint longitudinal models (L-MI and L-FB). For these two types of methods, we note that the credible intervals of Bayesian models are wider compared to the bootstrapped confidence intervals evaluated for the multiple imputation models. Numerical summary statistics for other CEA quantities (including incremental results) for the analysis of the PBS study are separately provided in Table~\ref{tabestpbs} in Appendix~\ref{app_pbs}.  

Figure~\ref{fig_PBS_cea} shows the CEACs associated with each method applied to the data from the PBS study.
\begin{figure}[!h]
\centering
\textbf{Cost-Effectiveness Acceptability Curve}\par
\includegraphics[scale=0.45]{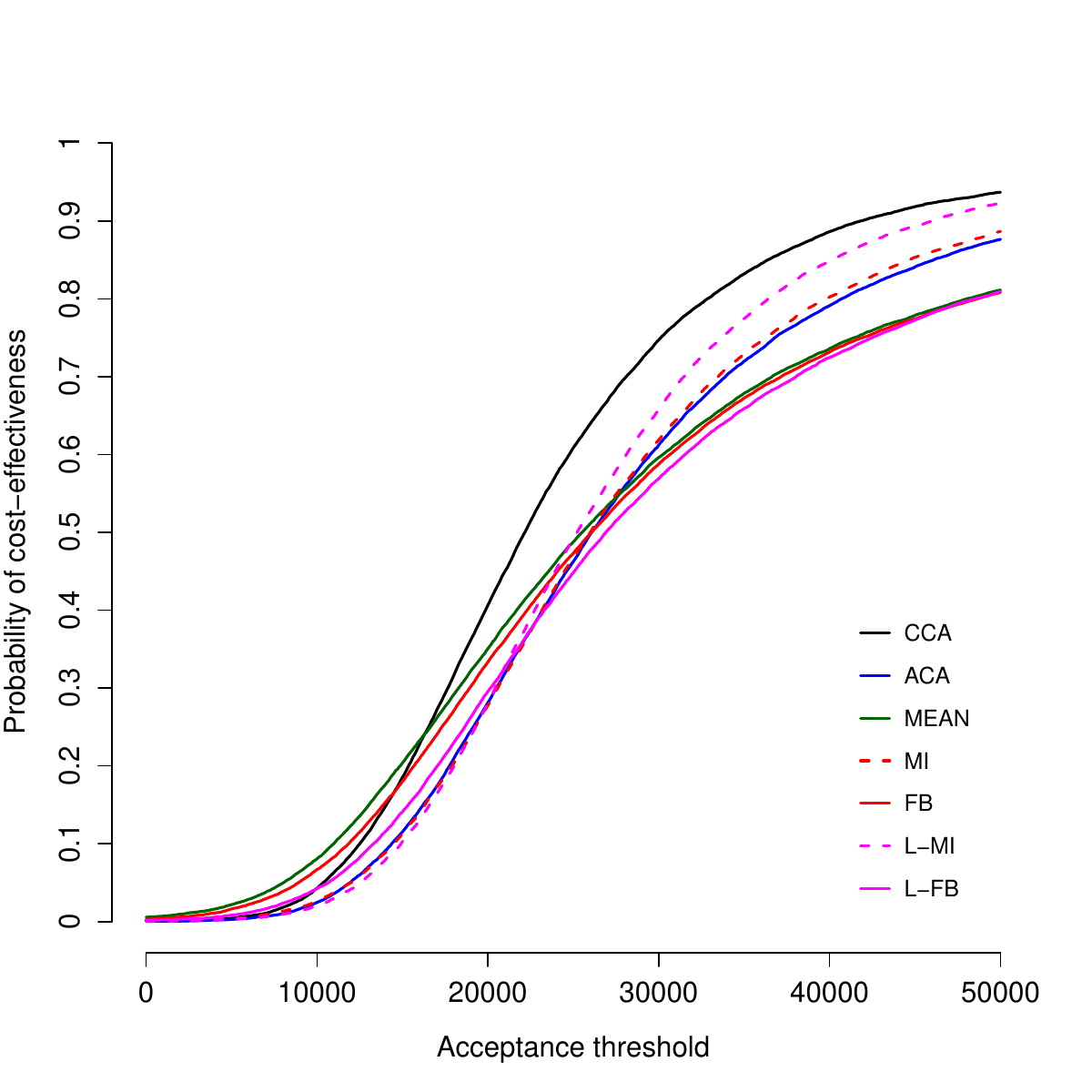}
\caption{Cost-effectiveness acceptability curves associated with seven alternative missingness approaches fitted to the PBS study. Solid and dashed lines are used to indicate the results associated with the models fitted using Bayesian (CCA, ACA, MEAN, FB and L-FB) and multiple imputation methods (MI and L-MI). A variation of the acceptance thresholds up to \pounds{50,000} is considered.}\label{fig_PBS_cea}
\end{figure}
The curves show a similar pattern across all threshold values and the approaches compared. The largest difference is observed for the results of CCA, which indicate an increased chance of cost-effectiveness with respect to the other methods (between $0.1$ and $0.2$). Results from joint aggregated models (MI and FB) and joint longitudinal models (L-MI and L-FB) do not show substantial differences, even though models fitted under a Bayesian framework show slightly lower curves with respect to those from multiple imputation methods for threshold values above $\pounds 30,000$.

\section{Discussion}\label{discussion}
The objective of this paper was to assess the impact that alternative missingness approaches may have on trial-based CEA results. Focus was given to the differences between joint models that properly account for the longitudinal nature of the utility and cost data and those that are fitted to the aggregated effectiveness and total cost variables. For comparison, we also included other approaches, such as case deletion and baseline imputation methods, which are routinely used by practitioners. The methods were fitted and their impact on the results compared using a brief simulation study and two real case studies. 

\subsection{Summary of the Results}\label{disca}
In the simulation study we assessed the performance of the methods across a range of scenarios. Results from these analyses suggest that, while under MCAR all methods perform well (thus favouring the use of simpler approaches), they show considerable differences between the two MAR mechanisms explored (MAR1 and MAR2). Overall, joint aggregate and longitudinal models are associated with the best performance and lead to similar results when missingness exclusively depends on the observed baseline values (MAR1). However, when missingness also depends on some observed follow-up values (MAR2), longitudinal models are the only approach which shows unbiased results. We note that these results are based on a limited number of scenarios which were defined based on a simplified setting. A more comprehensive assessment could be obtained by extending (e.g.~more types of mechanisms) the scenarios assessed which, for brevity, we did not consider here. In the application to the two case studies, the impact on CEA conclusions of the different methods varies. In the MenSS trial, CCA is the only method suggesting the severe lack of cost-effectiveness for the new intervention, while all other approaches indicate an opposite conclusion. Among these, joint aggregated models (MI and FB) show similar conclusions to simpler methods (ACA and MEAN), and suggest the most favourable results for the new intervention. Joint longitudinal models (L-MI and L-FB) indicate relatively milder conclusions. In the PBS trial, more favourable cost-effectiveness results are obtained under CCA, while all other methods indicate more moderate conclusions, even though variations are relatively small.

\subsection{Conclusions}\label{discb}
The results from our analyses provide interesting elements for a discussion. First, all aggregate methods may be associated with incorrect inferences even under MAR, since they do not incorporate all post-baseline values. Conversely, models that properly account for the longitudinal nature of the data are fitted using all the available observations and lead valid inferences under MAR. Second, the magnitude of the differences between the methods may vary depending on factors such as sample sizes and the proportions of missing values. For example, when the number of completers is relatively small (as in the MenSS study), differences between the two types of joint models are more likely to increase since aggregate models discard a considerable portion of post-baseline data. Conversely, when the number of completers is large (as in the PBS study), variations between the methods are likely to be small and the potential gain of using longitudinal models will be more limited. A quick strategy which can provide an idea about the potential benefit of using joint longitudinal models with respect to joint aggregated models (under MAR) is to compare the distributions of the utilities and costs between the completers and the non-completers at baseline and all follow-up points in the study. The presence of systematic differences between these distributions suggests that incorporating all the available data using a longitudinal approach may improve the estimation of the quantities of interest under MAR, with respect to models that, even partially, ignore this~information. 

Both MI and FB are valid approaches that can be used to fit joint longitudinal models, although other approaches could also be considered (e.g.~maximum likelihood methods). In our analyses, we did not find substantial differences in terms of the complexity of the implementation for the two methods. We note that, by virtue of the separation of the analysis and imputation steps, MI allows to tackle computational issues in a faster way compared to Bayesian methods, particularly when multiple partially-observed covariates are included in the analysis. However, we also note that, for the purpose of generating replications of parameter estimates (used in the cost-effectiveness assessment) different approaches are available to combine MI and bootstrap methods~\citep{Brand}. Even though recent studies compared the performance of alternative approaches and provided some recommendations~\citep{Schomaker}, no consensus has been reached yet about which one to use in general situations. This may be problematic in CEA, where the objective is decision-making rather than statistical inference per se~\citep{claxton1999}. Despite being computationally more challenging, Bayesian models naturally allow the simultaneous propagation of uncertainty throughout the model, which can be assessed in the posterior distributions of the parameters. On the other hand, when using MI, the analyst must ensure that: the analysis and imputation models are correctly specified to avoid biased results~\citep{van2018flexible}; the uncertainty in the estimates is generated via bootstrapping in a way that is consistent with both the analysis and imputation models. When the complexity of the model is increased to tackle different issues of the data (e.g.~clustering, correlation, non-normal distributions), then the task of ensuring that all the different components of an MI analysis are correctly specified may become extremely difficult compared to using a more flexible Bayesian approach.

A limitation of this study is that, given its focus on current practice, additional complexities that affect CEA data were not explicitly considered. Specifically, inferences were derived assuming separate normal distributions for the utility and cost data. These, however, are unlikely to be realistic modelling choices, since both types of data are characterised by different complexities that should be taken into account in the analysis. These includes: the potential correlation between the outcomes, the skewness and the presence of spikes at the boundaries in the empirical distributions~\citep{OHagan,Basu,Ng,Gabriob}. Finally, although MAR is the default assumption in current practice, it can never be checked from the data and inferences under MAR may be incorrect. Thus, it is extremely important that plausible MNAR departures are explored and their impact on the final conclusions is assessed in sensitivity analysis. Different approaches are available to fit the models under MNAR, which can be implemented using either a MI or FB approach~\citep{Daniels,VanBuuren,Carpenter,Mason,Leurent2018b,Gabriob}. 

In conclusion, aggregated models that ignore the longitudinal nature of the data may discard considerable proportions of post-baseline utility and cost values and potentially mislead the decision-making process. Conversely, joint longitudinal models take into account the longitudinal aspect and define the MAR assumption using all the available evidence in the study. In addition, when external information is available, these models can be extended to assess the robustness of the results to a range of plausible departures from MAR.

\section{Funding}
The MenSS trial was supported by a Health Technology Assessment (HTA) grant from the National Institute for Health Research. Ref. 10/131/01 \url{http://www.nets.nihr.ac.uk/projects/hta/1013101}. The PBS study was funded by the Health Technology Assessment program of the National Institute of Health Research (Reference 10/104/13).

\ack{The authors are grateful to two anonymous reviewers for careful reading and thoughtful comments that have greatly improved an earlier version of the paper. We would also like to acknowledge the hard work of all of the people involved in the MenSS and PBS trials and to thank them for providing us with access to their data.

Dr Andrea Gabrio is partially funded in his PhD programme at University College London by a research grant sponsored by The Foundation BLANCEFLOR Boncompagni Ludovisi, n\'{e}e Bildt.

Dr Gianluca Baio is partially supported as the recipient of an unrestricted research grant sponsored by Mapi Group at University College London.
}

\clearpage

\begin{appendices}


\section{Case~Studies}\label{A1}

\subsection{MenSS Trial}\label{app_menss}

\begin{table}[!h]
\centering
\scalebox{0.7}{
\begin{tabular}{ccccccc|c|c|c}
\toprule
 \multicolumn{7}{l|}{\textbf{Missing data patterns}} & \textbf{Control} ($n_1=75$)  & \textbf{Intervention} ($n_2=84$) & \textbf{Total} ($n=159$)\\[2ex]
 $u_0$ &  $u_1$ &  $c_1$ &  $u_2$ & $c_2$ & $u_3$ & $c_3$ & $n_{1}(\%)$ & $n_{2}(\%)$ & $n(\%)$\\
\midrule
  \cmark & \cmark & \cmark & \cmark & \cmark & \cmark & \cmark  & 27 (36\%) & 19(23\%) & 46(29\%)\\ [1ex]
 \xmark & \cmark & \cmark & \cmark & \cmark & \cmark & \cmark  &  0 (0\%) & 1(1\%) & 1(<1\%)\\ [1ex]
\cmark & \cmark & \cmark & \cmark & \cmark & \xmark & \xmark & 1 (1\%) & 0(0\%) & 1(<1\%)\\[1ex] 
 \cmark & \cmark & \cmark & \xmark & \xmark & \cmark & \cmark & 4 (6\%) & 1(1\%) & 5(3\%)\\[1ex] 
\cmark & \xmark & \xmark & \cmark & \cmark & \cmark & \cmark &  5 (4\%) & 3(3\%) & 8(5\%)\\ [1ex]
 \xmark & \cmark & \cmark & \xmark & \xmark & \cmark & \cmark &  1 (1\%) & 0(0\%) & 1(<1\%)\\ [1ex]
\cmark & \xmark & \xmark & \cmark & \cmark & \xmark & \xmark &   2(3\%) & 0(0\%) & 2(1\%)\\ [1ex]
  \cmark & \cmark & \cmark & \xmark & \xmark & \xmark & \xmark  & 1 (1\%) & 2(2\%) & 3(2\%)\\ [1ex]
 \cmark & \xmark & \xmark & \xmark & \xmark & \cmark & \cmark  &   4 (6\%) & 9(11\%) & 12(8\%)\\[1ex] 
 \xmark & \xmark & \xmark & \xmark & \xmark & \cmark & \cmark &  2 (3\%) & 3(3\%) & 5(3\%)\\ [1ex]
\cmark & \xmark & \xmark & \xmark & \xmark & \xmark & \xmark & 28 (38\%) & 38(46\%) & 66(42\%)\\[1ex] 
\xmark & \xmark & \xmark & \xmark & \xmark & \xmark & \xmark &  0 (0\%) & 8(10\%) & 8(5\%)\\ 
\bottomrule
\end{tabular}
}\caption{Missingness patterns for the utility and cost variables in the MenSS study. For each pattern and treatment group $t=(1,2)$, the number ($n_{t}$) and proportions of subjects are reported. We denote the presence and absence of values or individuals within each pattern with the symbols \cmark and \xmark, respectively.}\label{tabpatmenss}
\end{table}

\begin{figure}[!h]
\centering
\subfloat[control]{\includegraphics[scale=0.4]{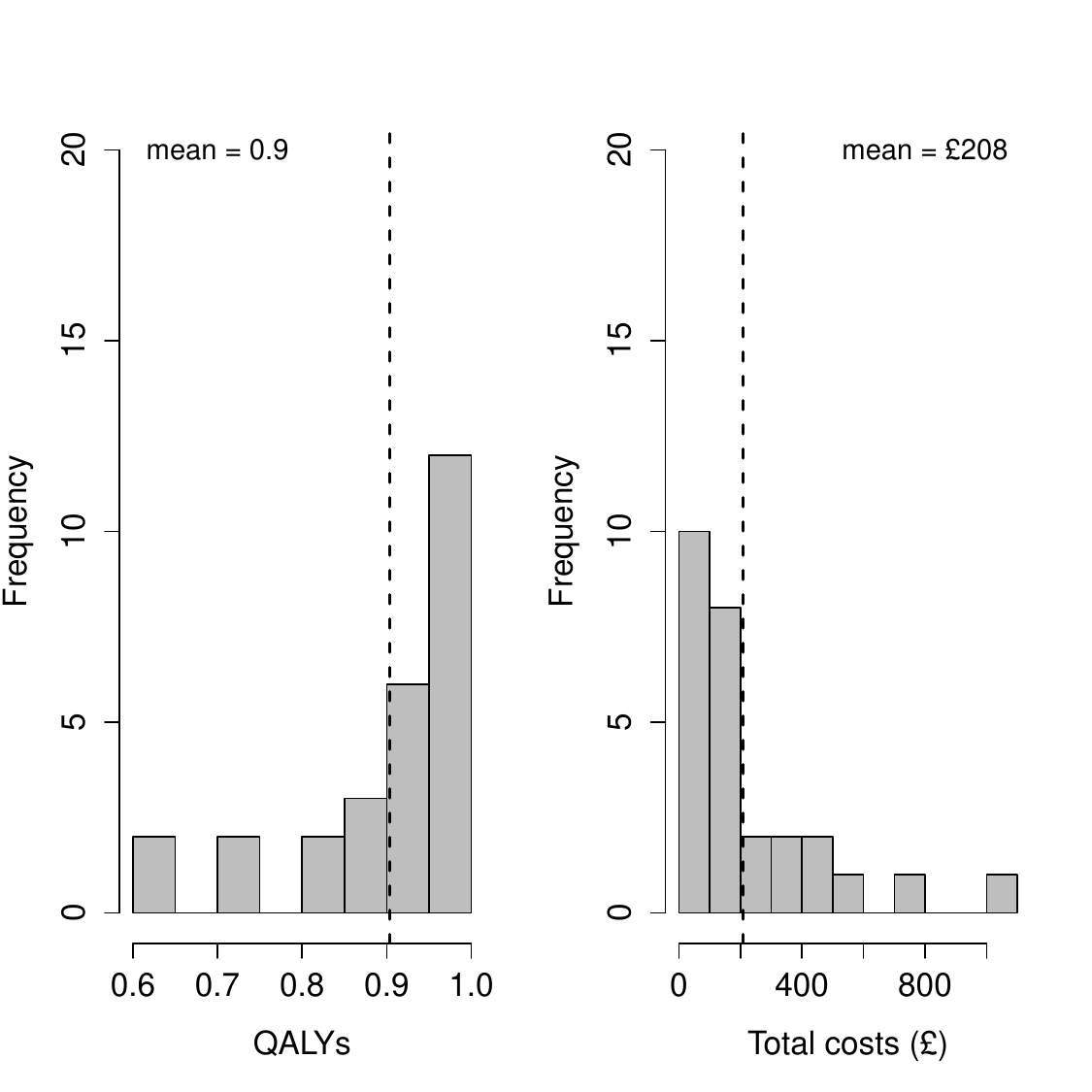}}
\subfloat[intervention]{\includegraphics[scale=0.4]{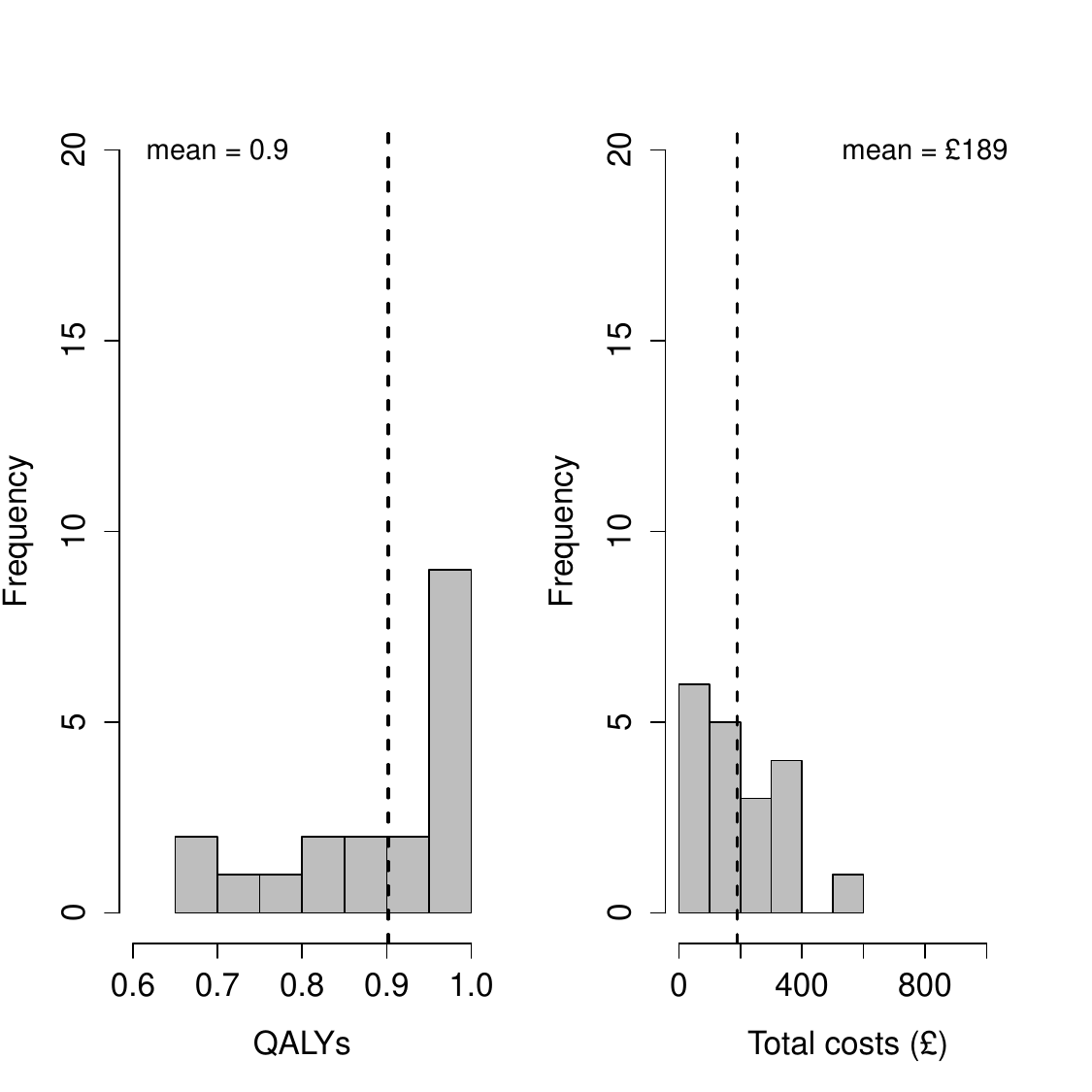}}
\caption{QALYs and total cost distributions for the control (panel a) and intervention (panel b) group in the MenSS trial. A dashed line is drawn in correspondence of the mean for each variable and the value is reported in the plots. Costs are expressed in \pounds.}
\label{data_Menss}
\end{figure}

\begin{table}[H]
\centering
\scalebox{0.7}{
\begin{tabular}{r|ccccccc}
  \toprule
 & CCA & ACA & MEAN & MI & FB & L-MI & L-FB \\ 
  \midrule
\textbf{Control} &  &  &  &  &  &  &  \\ [1ex]
   QALYs & 0.9 (0.88;0.93) & 0.87 (0.84;0.9) & 0.87 (0.84;0.9) & 0.87 (0.84;0.9) & 0.87 (0.84;0.9) & 0.86 (0.83;0.89) & 0.87 (0.83;0.9) \\ 
  Total costs & 207 (136;280) & 207 (136;280) & 208 (134;281) & 201 (123;287) & 208 (135;281) & 243 (146;337) & 233 (163;303) \\ [1ex]
 \textbf{Intervention} &  &  &  &  &  &  &  \\ [1ex]
   QALYs & 0.9 (0.87;0.93) & 0.91 (0.88;0.94) & 0.91 (0.88;0.95) & 0.92 (0.88;0.95) & 0.91 (0.88;0.95) & 0.88 (0.84;0.93) & 0.89 (0.86;0.93) \\ 
 Total costs & 189 (101;275) & 189 (101;275) & 189 (101;276) & 186 (122;253) & 189 (102;277) & 192 (114;263) & 179.85 (98;263) \\ [1ex]
 \textbf{Incremental} &  &  &  &  &  &  &  \\ [1ex]
  QALYs  & 0 (-0.05;0.05) & 0.04 (-0.01;0.09) & 0.05 (-0.01;0.1) & 0.04 (0;0.09) & 0.05 (-0.01;0.1) & 0.02 (-0.03;0.08) & 0.03 (-0.03;0.08) \\ 
 Total costs & -18 (-154;117) & -18 (-154;117) & -19 (-154;116) & -16 (-138;103) & -19 (-156;117) & -51 (-187;70) & -53 (-182;77) \\ [1ex]
  ICER & 9481 & -471 & -404 & -367 & -397 & -2181 & -2102 \\ 
   \bottomrule
\end{tabular}
}\caption{Estimates and $95\%$ credible/confidence intervals for key parameters in the economic analysis of the MenSS trial. Results are shown for: mean QALYs and total costs in the control and intervention group; incremental QALYs and total cost means, and ICER, between the two interventions. For each quantity, estimates associated with seven alternative approaches are displayed: complete and available case analysis (CCA and ACA), mean baseline imputation (MEAN), joint aggregated and longitudinal models, either fitted in a frequentist (MI and L-MI) or Bayesian framework (FB and L-FB).}\label{tabestmenss}
\end{table}

\subsection{PBS Trial}\label{app_pbs}

\begin{table}[!h]
\centering
\scalebox{0.7}{
\begin{tabular}{cccccc|c|c|c}
\toprule
 \multicolumn{6}{l|}{\textbf{Missing data patterns}} & \textbf{Control} ($n_1=136$)  & \textbf{Intervention} ($n_2=108$) & \textbf{Total} ($n=244$)\\[2ex]
 $u_0$ &  $c_0$ &  $u_1$ &  $c_2$ & $u_2$ & $c_2$ & $n_{1}(\%)$ & $n_{2}(\%)$ & $n(\%)$\\
\midrule
\cmark & \cmark & \cmark & \cmark & \cmark & \cmark   & 108 (80\%) & 96(89\%) & 204(84\%)\\ [1ex]
\xmark & \cmark & \cmark & \cmark & \cmark & \cmark   &  7 (5\%) & 5(5\%) & 12(5\%)\\ [1ex]
\cmark & \cmark & \xmark & \cmark & \cmark & \cmark   & 4 (3\%) & 1(1\%) & 5(2\%)\\[1ex] 
 \cmark & \cmark & \cmark & \cmark & \xmark & \cmark & 2 (1\%) & 1(1\%) & 3(1\%)\\[1ex] 
 \cmark & \cmark & \xmark & \xmark & \cmark & \cmark   &  4 (3\%) & 1(1\%) & 5(2\%)\\ [1ex]
\cmark & \cmark & \xmark & \xmark & \xmark & \xmark  &  4 (4\%) & 4(3\%) & 8(3\%)\\ [1ex]
\xmark & \cmark & \xmark & \cmark & \cmark & \cmark  &   2(1\%) & 0(0\%) & 2(1\%)\\ [1ex]
 \cmark & \cmark & \cmark & \cmark & \xmark & \xmark   & 2 (1\%) & 0(0\%) & 2(1\%)\\ [1ex]
\cmark & \cmark & \xmark & \cmark & \xmark & \cmark   &   3 (2\%) & 0(0\%) & 3(1\%)\\
\bottomrule
\end{tabular}
}\caption{Missingness patterns for the utility and cost variables in the PBS study. For each pattern and treatment group $t=(1,2)$, the number ($n_{t}$) and proportions of subjects are reported. We denote the presence and absence of values or individuals within each pattern with the symbols \cmark and \xmark, respectively.}\label{tabpatpbs}
\end{table}

\begin{figure}[!h]
\centering
\subfloat[control]{\includegraphics[scale=0.4]{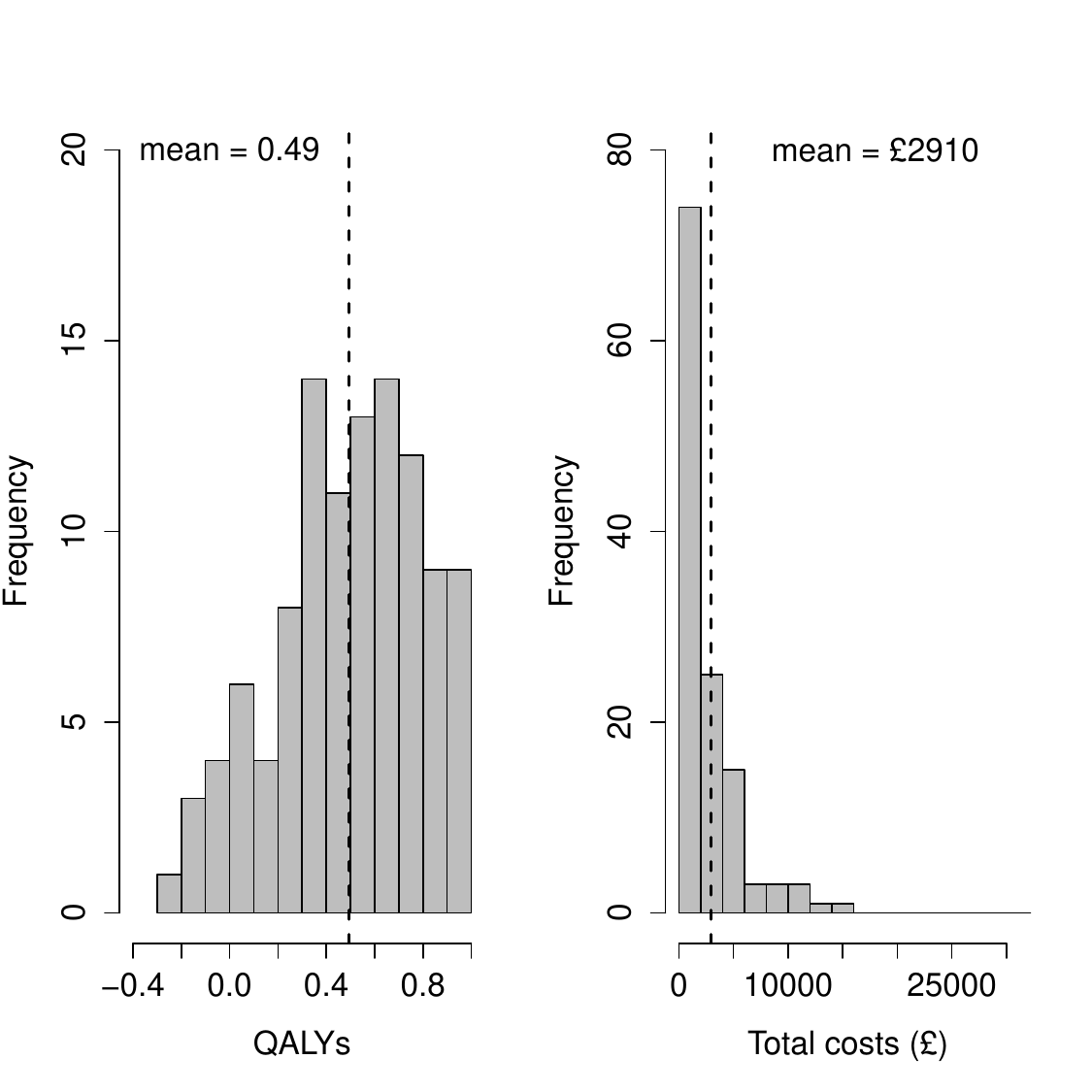}}
\subfloat[intervention]{\includegraphics[scale=0.4]{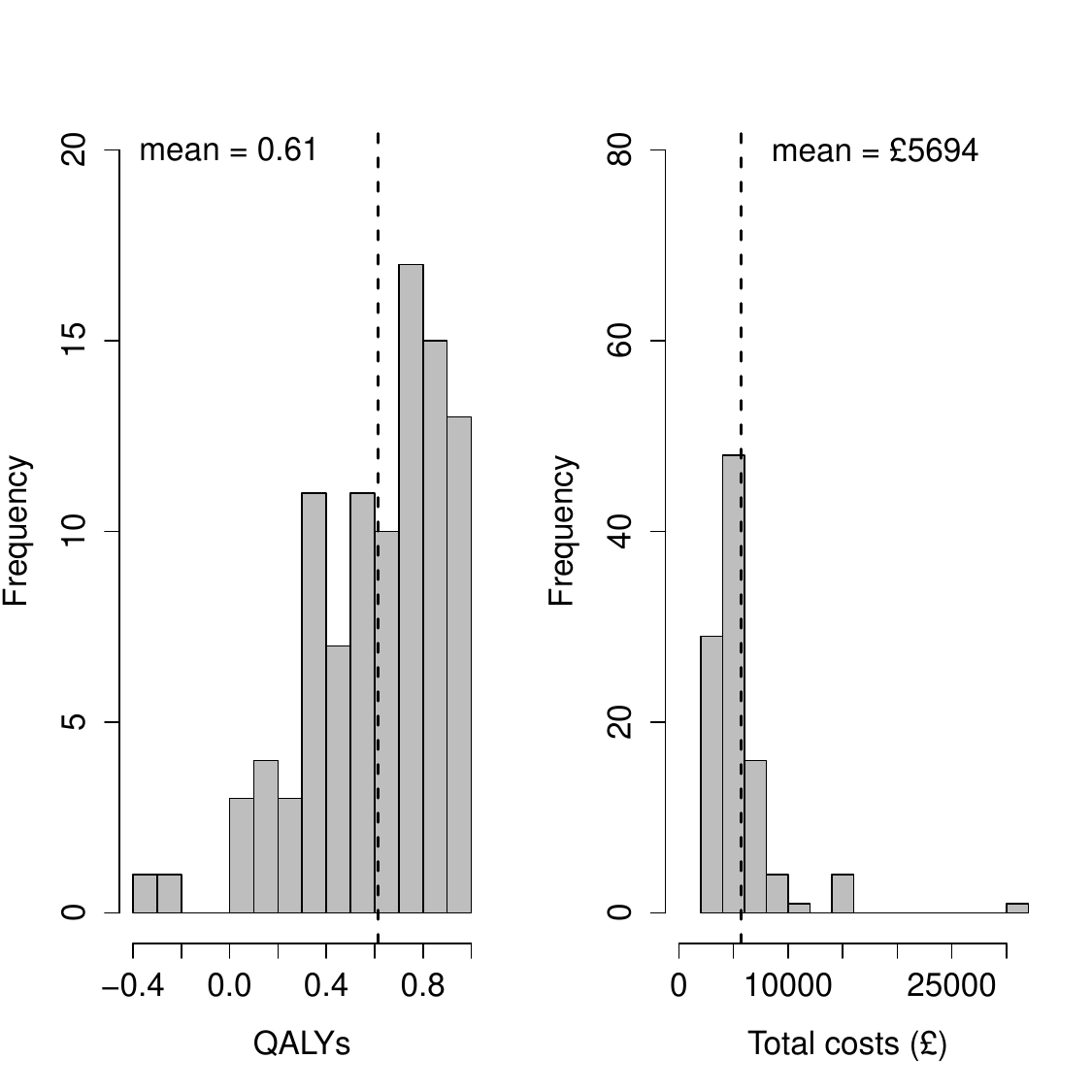}}
\caption{QALYs and total cost distributions for the control (panel a) and intervention (panel b) group in the PBS trial. A dashed line is drawn in correspondence of the mean for each variable and the value is reported in the plots. Costs are expressed in \pounds.}
\label{data_PBS}
\end{figure}

\begin{table}[!h]
\centering
\scalebox{0.7}{
\begin{tabular}{r|ccccccc}
  \toprule
 & CCA & ACA & MEAN & MI & FB & L-MI & L-FB \\ 
  \midrule
\textbf{Control} &  &  &  &  &  &  &  \\ [1ex]
  QALYs & 0.49 (0.46;0.53) & 0.5 (0.46;0.54) & 0.51 (0.47;0.55) & 0.51 (0.46;0.54) & 0.51 (0.47;0.56) & 0.51 (0.46;0.53) & 0.51 (0.46;0.55) \\ 
  Total costs & 2967 (2071;3834) & 2966 (2070;3833) & 3133 (2222;4028) & 3129 (2533;3925) & 3208 (2373;4034) & 3298 (2520;3891) & 3061 (2223;3914) \\ [1ex]
 \textbf{Intervention} &  &  &  &  &  &  &  \\ [1ex]
  QALYs & 0.61 (0.56;0.67) & 0.61 (0.55;0.66) & 0.59 (0.53;0.64) & 0.59 (0.56;0.64) & 0.59 (0.53;0.65) & 0.59 (0.56;0.64) & 0.59 (0.53;0.65) \\ 
   Total costs & 5655 (4796;6497) & 5716 (4849;6564) & 5068 (4265;5870) & 5105 (4626;6173) & 5209 (4388;6011) & 5357 (4652;6181) & 5274 (4539;5994) \\ [1ex]
 \textbf{Incremental} &  &  &  &  &  &  &  \\ [1ex]
  QALYs & 0.12 (0.04;0.21) & 0.11 (0.02;0.19) & 0.08 (-0.01;0.16) & 0.08 (0.02;0.13) & 0.08 (0;0.16) & 0.08 (0.03;0.13) & 0.08 (0;0.16) \\ 
   Total costs & 2688 (1243;4167) & 2750 (1294;4237) & 1935 (473;3421) & 1976 (880;2947) & 2001 (589;3365) &  2059 (976;2979) & 2213 (862;3504) \\ [1ex]
  ICER & 22123 & 26069 & 25483 & 25927 & 25969 & 25097 & 27026 \\ 
   \bottomrule
\end{tabular}
}\caption{Estimates and $95\%$ credible/confidence intervals for key parameters in the economic analysis of the PBS trial. Results are shown for: mean QALYs and total costs in the control and intervention group; incremental QALYs and total cost means, and ICER, between the two interventions. For each quantity, estimates associated with seven alternative approaches are displayed: complete and available case analysis (CCA and ACA), mean baseline imputation (MEAN), joint aggregated and longitudinal models, either fitted in a frequentist (MI and L-MI) or Bayesian framework (FB and L-FB).}\label{tabestpbs}
\end{table}

\end{appendices}

\bibliographystyle{apa}
\bibliography{missing_methods}

\begin{thebibliography}{}

\bibitem[\protect\astroncite{Bailey et~al.}{2016}]{Hunterb}
Bailey, J., Webster, R., Hunter, R., Griffin, M., N., F., Rait, G., Estcourt,
  C., Michie, S., Anderson, J., Stephenson, J., Gerressu, M., Sinag~Ang, C.,
  and Murray, E. (2016).
\newblock The men's safer sex project: intervention development and feasibility
  randomised controlled trial of an interactive digital intervention to
  increase condom use in men.
\newblock {\em Health Technology Assessment}, 20.

\bibitem[\protect\astroncite{Basu and Manca}{2012}]{Basu}
Basu, A. and Manca, A. (2012).
\newblock Regression estimators for generic health-related quality of life and
  quality-adjusted life years.
\newblock {\em Medical Decision Making}, 1:56--69.

\bibitem[\protect\astroncite{Brand et~al.}{2018}]{Brand}
Brand, J., van Buuren, S., le~Cessie, S., and van~den Hout, W. (2018).
\newblock Combining multiple imputation and bootstrap in the analysis of
  cost-effectiveness trial data.
\newblock {\em Statistics in Medicine}, 38:210--220.

\bibitem[\protect\astroncite{Brooks et~al.}{2011}]{Brooks}
Brooks, S., Gelman, A., Jones, G., and Meng, X. (2011).
\newblock {\em Handbook of {Markov Chain Monte Carlo}}.
\newblock CRC press.

\bibitem[\protect\astroncite{Carpenter and Kenward}{2012}]{Carpenter}
Carpenter, G. and Kenward, M. (2012).
\newblock {\em Multiple Imputation and Its Applications}.
\newblock John Wiley and Sons.

\bibitem[\protect\astroncite{Claxton}{1999}]{claxton1999}
Claxton, K. (1999).
\newblock The irrelevance of inference: a decision-making approach to the
  stochastic evaluation of health care technologies.
\newblock {\em Journal of Health Economics}, 18:341--364.

\bibitem[\protect\astroncite{Daniels and Hogan}{2008}]{Daniels}
Daniels, M. and Hogan, J. (2008).
\newblock {\em Missing Data in Longitudinal Studies: Strategies for Bayesian
  Modeling and Sensitivity Analysis}.
\newblock Chapman and Hall, New York.

\bibitem[\protect\astroncite{Drummond et~al.}{2005}]{Drummond}
Drummond, M., Schulpher, M., Claxton, K., Stoddart, G., and Torrance, G.
  (2005).
\newblock {\em Methods for the economic evaluation of health care programmes.
  3rd ed}.
\newblock Oxford university press, Oxford.

\bibitem[\protect\astroncite{Efron}{1981}]{Efron}
Efron, B. (1981).
\newblock Nonparametric standard errors and confidence intervals.
\newblock {\em Canad. J. Statist.}, 9:139--172.

\bibitem[\protect\astroncite{Gabrio et~al.}{2019a}]{gabrio2019bayesian}
Gabrio, A., Daniels, M.~J., and Baio, G. (2019a).
\newblock A bayesian parametric approach to handle missing longitudinal outcome
  data in trial-based health economic evaluations.
\newblock {\em Journal of the Royal Statistical Society: Series A (Statistics
  in Society)}.

\bibitem[\protect\astroncite{Gabrio et~al.}{2017}]{Gabrio}
Gabrio, A., Mason, A., and Baio, G. (2017).
\newblock Handling missing data in within-trial cost-effectiveness analysis: A
  review with future recommendations.
\newblock {\em PharmacoEconomics-Open}, 1:79--97.

\bibitem[\protect\astroncite{Gabrio et~al.}{2019b}]{Gabriob}
Gabrio, A., Mason, A., and Baio, G. (2019b).
\newblock A full {Bayesian} model to handle structural ones and missingness in
  economic evaluations from individual‐level data.
\newblock {\em Statistics in Medicine}, 38:1399--1420.

\bibitem[\protect\astroncite{Gelman et~al.}{2004}]{Gelman2}
Gelman, A., Carlin, J., Stern, H., and Rubin, D. (2004).
\newblock {\em Bayesian Data Analysis - 2nd edition}.
\newblock Chapman and Hall, New York, NY.

\bibitem[\protect\astroncite{Gomes et~al.}{2019}]{gomes2019copula}
Gomes, M., Radice, R., Camarena~Brenes, J., and Marra, G. (2019).
\newblock Copula selection models for non-gaussian outcomes that are missing
  not at random.
\newblock {\em Statistics in medicine}, 38(3):480--496.

\bibitem[\protect\astroncite{Hassiotis et~al.}{2018}]{Hassiotis}
Hassiotis, A., Poppe, M., Strydom, A., Vickerstaff, V., Hall, I., Crabtree, J.,
  Omar, R., King, M., Hunter, R., Biswas, A., Cooper, V., Howie, W., and
  Crawford, M. (2018).
\newblock Clinical outcomes of staff training in positive behaviour support to
  reduce challenging behaviour in adults with intellectual disability: cluster
  randomised controlled trial.
\newblock {\em The British Journal of Psychiatry}, 212:161--168.

\bibitem[\protect\astroncite{Leurent et~al.}{2018a}]{Leurent2018}
Leurent, B., Gomes, M., and Carpenter, J. (2018a).
\newblock Missing data in trial-based cost-effectiveness analysis: An
  incomplete journey.
\newblock {\em Health Economics}.

\bibitem[\protect\astroncite{Leurent et~al.}{2018b}]{Leurent2018b}
Leurent, B., Gomes, M., Faria, R., Morris, S., Grieve, R., and Carpenter, J.
  (2018b).
\newblock Sensitivity analysis for not-at-random missing data in trial-based
  cost-effectiveness analysis: A tutorial.
\newblock {\em PharmacoEconomics}, pages 1--13.

\bibitem[\protect\astroncite{Little}{1992}]{Little1992}
Little, R. (1992).
\newblock Regression with missing x's: A review.
\newblock {\em Journal of the American Statistical Association}, 87:1227--1237.

\bibitem[\protect\astroncite{Little and Rubin}{2002}]{Little2002}
Little, R. and Rubin, D. (2002).
\newblock {\em Statistical analysis with missing data}.

\bibitem[\protect\astroncite{Little}{1995}]{little1995modeling}
Little, R.~J. (1995).
\newblock Modeling the drop-out mechanism in repeated-measures studies.
\newblock {\em Journal of the american statistical association},
  90(431):1112--1121.

\bibitem[\protect\astroncite{Mason et~al.}{2012}]{Mason}
Mason, A., Richardson, S., Plewis, I., and Best, N. (2012).
\newblock Strategy for modelling nonrandom missing data mechanisms in
  observational studies using {Bayesian} methods.
\newblock {\em Journal of Official Statistics}, 28:279--302.

\bibitem[\protect\astroncite{Mason et~al.}{2018}]{Mason2018}
Mason, A.~J., Gomes, M., Grieve, R., and Carpenter, J.~R. (2018).
\newblock A bayesian framework for health economic evaluation in studies with
  missing data.
\newblock {\em Health economics}, 27(11):1670--1683.

\bibitem[\protect\astroncite{Ng et~al.}{2016}]{Ng}
Ng, E., Diaz-Ordaz, K., Grieve, R., Nixon, R., Thompson, S., and Carpenter, J.
  (2016).
\newblock Multilevel models for cost-effectiveness analyses that use cluster
  randomised trial data: An approach to model choice.
\newblock {\em Statistical Methods in Medical Research}, 25:2036--2052.

\bibitem[\protect\astroncite{Nixon and Thompson}{2005}]{Nixon}
Nixon, R. and Thompson, S. (2005).
\newblock Methods for incorporating covariate adjustment, subgroup analysis and
  between-centre differences into cost-effectiveness evaluations.
\newblock {\em Health Economics}, 14:1217--1229.

\bibitem[\protect\astroncite{Noble et~al.}{2012}]{Noble2012}
Noble, S., Hollingworth, W., and Tilling, K. (2012).
\newblock Missing data in trial-based cost-effectiveness analysis: the current
  state of play.
\newblock {\em Health Economics}, 21:187--200.

\bibitem[\protect\astroncite{O'Hagan and Stevens}{2001}]{OHagan}
O'Hagan, A. and Stevens, J. (2001).
\newblock A framework for cost-effectiveness analysis from clinical trial data.
\newblock {\em Health Economics}, 10:303--315.

\bibitem[\protect\astroncite{Plummer}{2010}]{Plummer}
Plummer, M. (2010).
\newblock {JAGS: Just Another Gibbs Sampler}.
\newblock \url{http://www-fis.iarc.fr/~martyn/software/jags/}.

\bibitem[\protect\astroncite{Rubin}{1987}]{Rubina}
Rubin, D. (1987).
\newblock {\em Multiple Imputation for Nonresponse in Surveys}.
\newblock John Wiley and Sons, New York,USA.

\bibitem[\protect\astroncite{Schomaker and Heumann}{2018}]{Schomaker}
Schomaker, M. and Heumann, C. (2018).
\newblock Bootstrap inference when using multiple imputation.
\newblock {\em Statistics in Medicine}, 37:2252--2266.

\bibitem[\protect\astroncite{Su and Yajima}{2015}]{Su}
Su, Y. and Yajima, M. (2015).
\newblock {Package `R2jags'}.
\newblock \url{https://cran.r-project.org/web/packages/R2jags/}.

\bibitem[\protect\astroncite{Sullivan et~al.}{2016}]{Sullivan2016}
Sullivan, T., White, I., Salter, A., Ryan, P., and Lee, K. (2016).
\newblock Should multiple imputation be themethod of choice for handling
  missing data in randomized trials?
\newblock {\em Statistical Methods in Medical Research}, 27:2610--2626.

\bibitem[\protect\astroncite{Van~Buuren}{2018}]{van2018flexible}
Van~Buuren, S. (2018).
\newblock {\em Flexible imputation of missing data}.
\newblock Chapman and Hall/CRC.

\bibitem[\protect\astroncite{Van~Buuren and
  Groothuis-Oudshoorn}{2011}]{VanBuuren}
Van~Buuren, S. and Groothuis-Oudshoorn, K. (2011).
\newblock mice: Multivariate imputation by chained equations in {R}.
\newblock {\em Journal of Statistical Software}, 45:1--67.

\bibitem[\protect\astroncite{van Buuren and
  Groothuis-Oudshoorn}{2019}]{micepkg}
van Buuren, S. and Groothuis-Oudshoorn, K. (2019).
\newblock {Package `mice'}.
\newblock \url{https://cran.r-project.org/web/packages/mice/}.

\bibitem[\protect\astroncite{Van~Hout et~al.}{1994}]{VanHout}
Van~Hout, B., Al, M., Gordon, G., Rutten, F., and Kuntz, K. (1994).
\newblock Costs, effects and c/e-ratios alongside a clinical trial.
\newblock {\em Health Economics}, 3:309--319.

\bibitem[\protect\astroncite{Van~Reenen and Oppe}{2015}]{EQ5D}
Van~Reenen, M. and Oppe, M. (2015).
\newblock {EQ-5D-3L User Guide Basic information on how to use the EQ-5D-3L
  instrument}.
\newblock
  \url{https://euroqol.org/wp-content/uploads/2016/09/EQ-5D-3L_UserGuide_2015.pdf}.

\bibitem[\protect\astroncite{White and Thompson}{2005}]{White2005}
White, I. and Thompson, S. (2005).
\newblock Adjusting for partially missing baseline measurements in randomized
  trials.
\newblock {\em Statistics in Medicine}, 24:993--1007.

\end{thebibliography}

\end{document}